\documentclass[acmtog, nonacm]{acmart} 

\usepackage{booktabs}
\usepackage{dcolumn}
\usepackage{array}
\usepackage{verbatim}
\usepackage{xcolor}
\usepackage{cleveref}
\usepackage[ruled]{algorithm2e}
\usepackage{bm}

\SetAlFnt{\small}
\SetAlCapFnt{\small}
\SetAlCapNameFnt{\small}
\SetAlCapHSkip{0pt}
\usepackage{parskip}

\acmJournal{TOG}

\usepackage{listings}
\usepackage{cleveref}
\usepackage{wrapfig}
\usepackage{subcaption}
\usepackage{wrapfig}
\usepackage{float}

\usepackage{dsfont}

\definecolor{derekBlue}{RGB}{144,210,236}
\definecolor{derekTableBlue}{RGB}{189,235,252}
\definecolor{iglGreen}{RGB}{153,203,67}
\definecolor{coralRed}{RGB}{250,114,104}
\definecolor{gray}{RGB}{180,180,180}
\definecolor{orange}{RGB}{255,165,0}
\definecolor{TechnionBlue}{RGB}{0,33,71}
\definecolor{lightgray}{gray}{0.65}

\definecolor{codered}{rgb}{0.8, 0.1, 0.05}
\definecolor{codegreen}{rgb}{0,0.35,0.5}
\definecolor{codeblue}{rgb}{0.2,0.45,0.65}
\definecolor{codepaleblue}{rgb}{0.15,0.4,0.15}
\definecolor{codeyellow}{rgb}{0.3,0.4,0.6}
\definecolor{codegray}{rgb}{0.3,0.3,0.3}
\definecolor{codepurple}{rgb}{0.58,0,0.82}
\definecolor{codepalegreen}{rgb}{0.4,0.6,0.3}
\definecolor{backcolor}{rgb}{0.99,0.98,0.97}
\definecolor{keywordcolor}{rgb}{0.75,0.35,0.15}
\definecolor{dslkeyword}{rgb}{0.65,0.35,0.6}
\lstdefinestyle{mystyle}{
    backgroundcolor=\color{backcolor},
    basicstyle=\ttfamily\footnotesize,
    commentstyle=\color{codegreen}\slshape,
    keywordstyle=\color{keywordcolor}\bfseries,
    numberstyle=\tiny\color{codegray},
    stringstyle=\color{codepaleblue},
    breakatwhitespace=false,      
    captionpos=b,
    keepspaces=true,
    numbers=left,
    numbersep=1.5em,
    showspaces=false,                
    rulecolor=\color{codegray},
    showstringspaces=false,
    showtabs=false,
    tabsize=2,
    frame=trlb,
    frameround = tttt,
    framerule=0em,
    xleftmargin=0.7em,
    xrightmargin=0.7em,
    emph={[2]distinct,def,true,false,return,if,for,goto,continue,switch,while,break,struct,class,do},
    emphstyle={[2]\color{keywordcolor}\bfseries},
    emph={[1]Interpolate, MCMCWeights,sort,Ones,RandInt,SamplePathAtPixel,ShapeAt},
    emphstyle={[1]\color{dslkeyword}\bfseries},
    escapeinside={\%*}{*)},
}
\newif\ifsubmit
\submitfalse
\definecolor{revcolor}{rgb}{0.1,0.7,0.9}
\definecolor{revcolor2}{rgb}{1.0,0.6,0.0}
\definecolor{warningcolor}{rgb}{1.0,0.0,0.0}

\ifsubmit
    \newcommand{\xiaochun}[1]{{}}

    \newcommand{\revbegin}[1]{} 
    \newcommand{\revend}[1]{} 
\else
    \newcommand{\xiaochun}[1]{{\bf\color{iglGreen}{Xiaochun: #1}}}
    
    \newcommand{\toshiya}[1]{}

    \newcommand{\revbegin}[1]{} 
    \newcommand{\revend}[1]{} 
\fi

\setcopyright{none}
\renewcommand\footnotetextcopyrightpermission[1]{}
\begin{document}
\title{Precomputed Lens Transport Maps}

\author{Yang Chen}
\authornote{All authors contributed equally to this work.}
\email{yangchen.jack@outlook.com}
\affiliation{%
  \institution{University of Waterloo}
  \country{Canada}
}

\author{Xiaochun Tong}
\authornotemark[1]
\email{xtong@uwaterloo.ca}
\affiliation{%
  \institution{University of Waterloo}
  \country{Canada}
}

\author{Afet Abzar}
\authornotemark[1]
\email{afet902813@gmail.com}
\affiliation{%
  \institution{University of Waterloo}
  \country{Canada}
}

\author{Leo Hanxu}
\authornotemark[1]
\email{leo.hanxu@uwaterloo.ca}
\affiliation{%
  \institution{University of Waterloo}
  \country{Canada}
}

\author{Matthew Avolio}
\authornotemark[1]
\email{mavolio@uwaterloo.ca}
\affiliation{%
  \institution{University of Waterloo}
  \country{Canada}
}

\author{Toshiya Hachisuka}
\email{thachisuka@uwaterloo.ca}
\affiliation{%
  \institution{University of Waterloo}
  \country{Canada}
}

\begin{abstract}
Accurate real-time simulation of lens optics remains challenging due to the computational expense of full ray tracing and the limitations of existing approximations. The commonly used pinhole model and thin-lens model ignore many optical effects seen in real-world lens systems such as distortion and chromatic aberration. Prior polynomial models approximate a mapping between incident rays and exitant rays through a lens system per wavelength. Prior neural models improve the accuracy of this mapping and also capture wavelength-dependent variations (e.g., chromatic aberration) by integrating wavelength as an input to a unified neural network. Common to those prior models is that they omit Fresnel intensity throughput, precluding accurate simulation of internal reflections and lens flares. We introduce a precomputed lens model that combines wavelength-aware inputs with Fresnel intensity outputs. By classifying rays as valid or occluded via a binary mask in a factorized representation, our method focuses regression on unblocked rays, improving accuracy near discontinuities. Our model avoids per-wavelength approximations in polynomial models and explicitly predicts Fresnel coefficients to enable accurate lens simulation. Designed for static, rotationally symmetric systems under geometric optics, our model captures various lens effects such as chromatic aberration, coma, and lens flares. Our method achieves improved accuracy over polynomial baselines and is \emph{an order of magnitude} faster than brute force ray tracing. Our method serves as a practical and scalable approach for simulating complex lens systems in applications requiring both accuracy and computational efficiency.
\end{abstract}

\begin{teaserfigure}
    \begin{minipage}{0.42\textwidth}
        \centering
        \includegraphics[width=\linewidth]{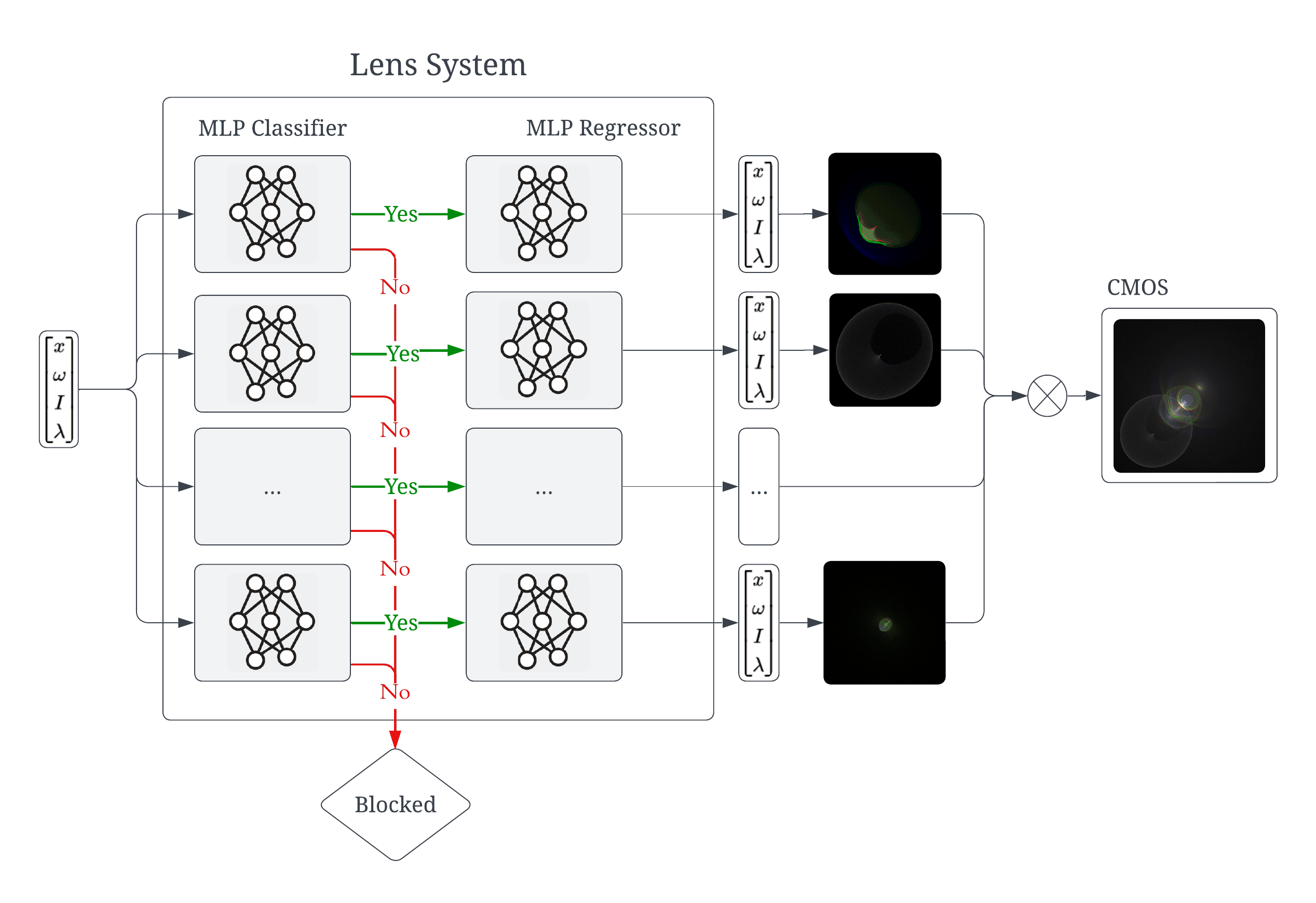}
    \end{minipage}%
    \hfill
    \begin{minipage}{0.24\textwidth}
        \begin{tabular}{ccc}
            \rotatebox[origin=c]{90}{59mm}&
            \begin{minipage}{0.33\linewidth}
                \includegraphics[width=\linewidth]{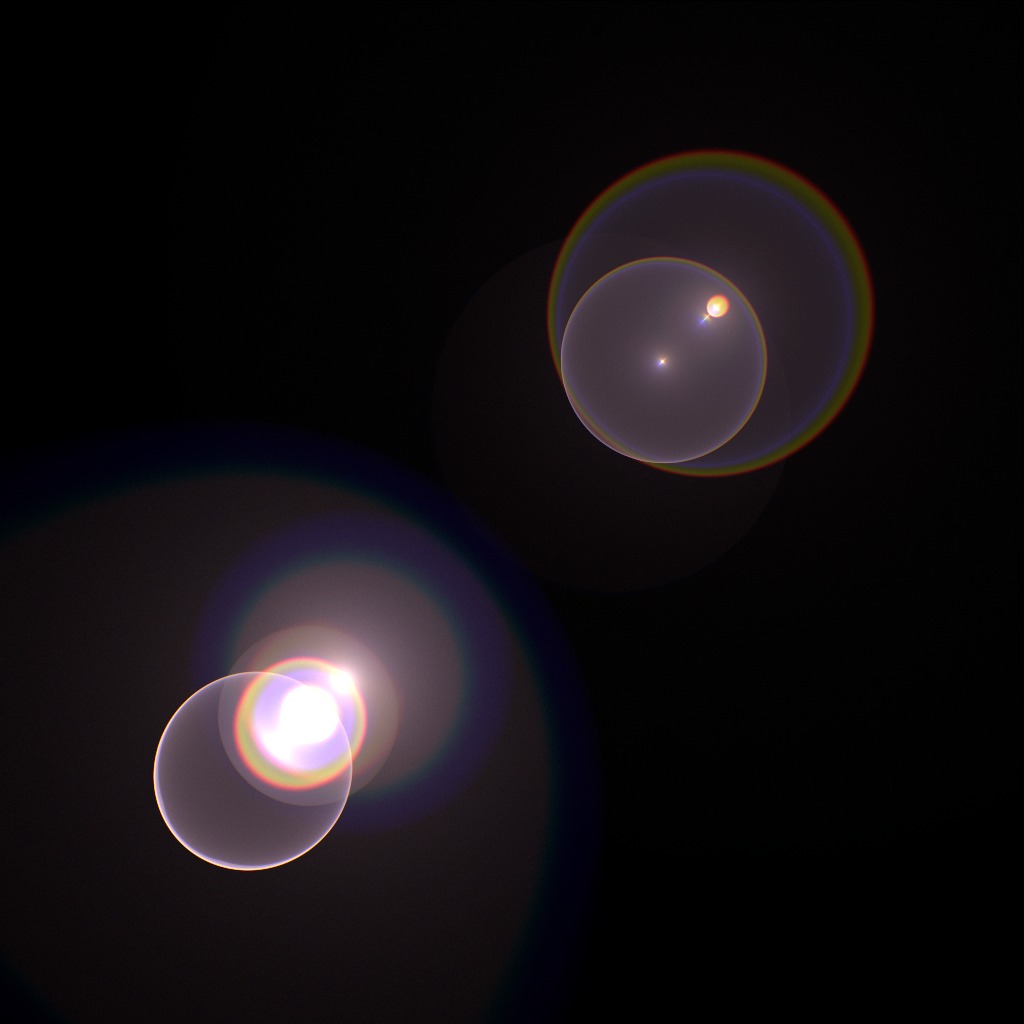}
            \end{minipage}&
            \begin{minipage}{0.33\linewidth}
                \includegraphics[width=\linewidth]{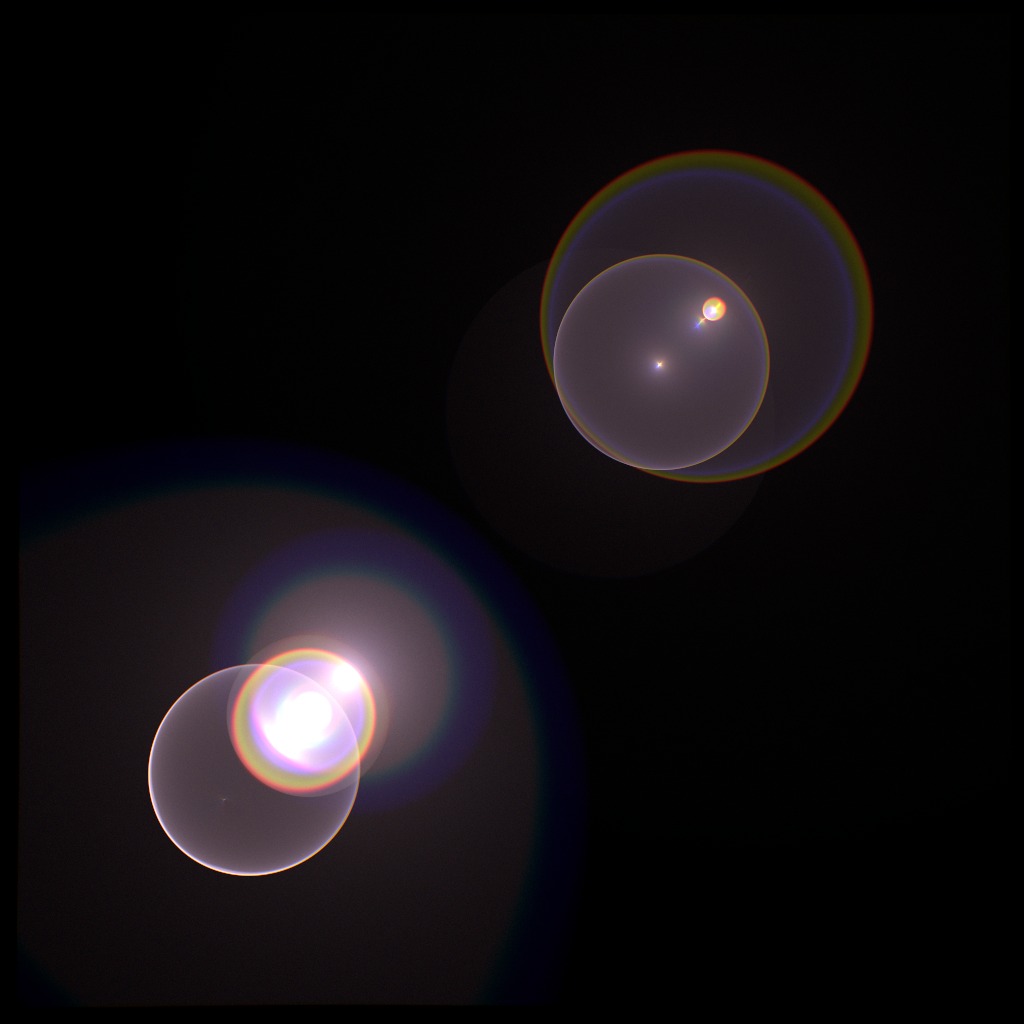}
            \end{minipage}\\
            & RT & Ours \\
            \rotatebox[origin=c]{90}{22mm}&
            \begin{minipage}{0.33\linewidth}
                \includegraphics[width=\linewidth]{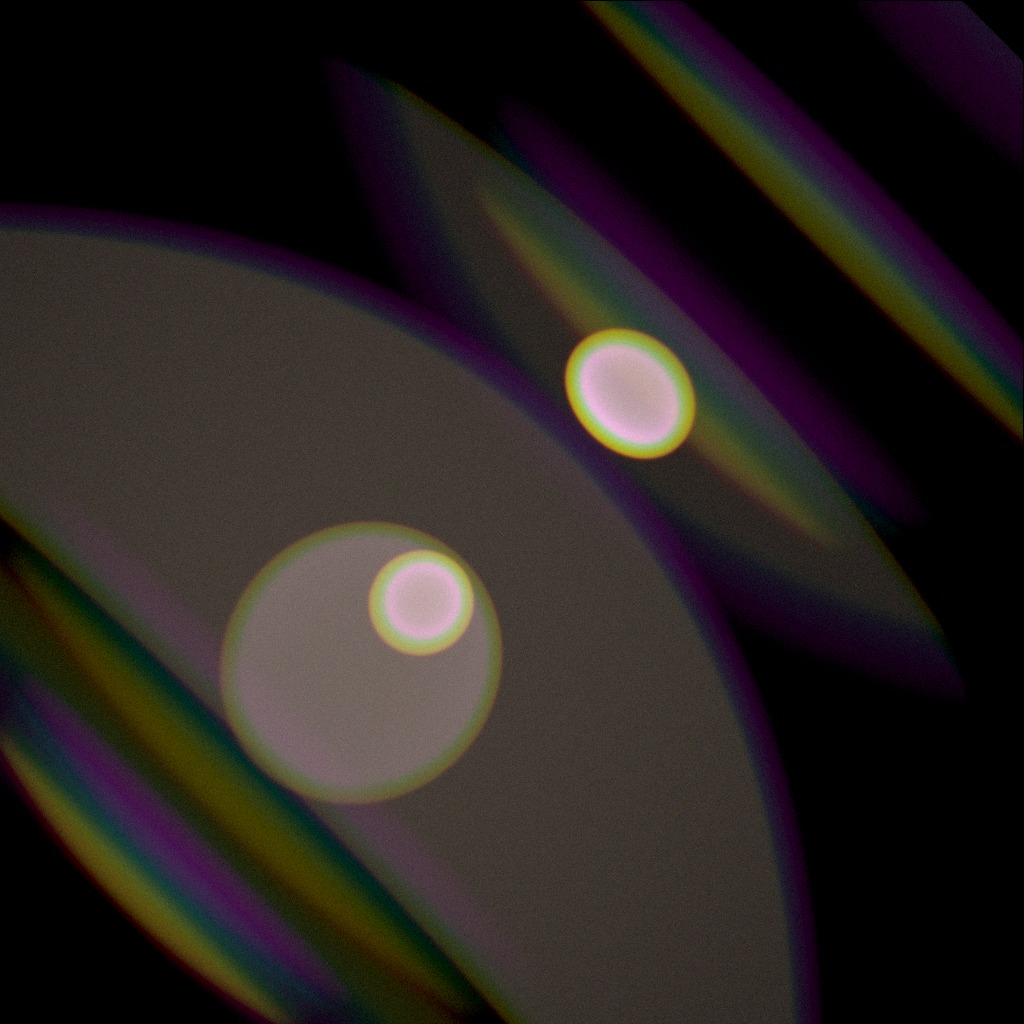}
            \end{minipage}&
            \begin{minipage}{0.33\linewidth}
                \includegraphics[width=\linewidth]{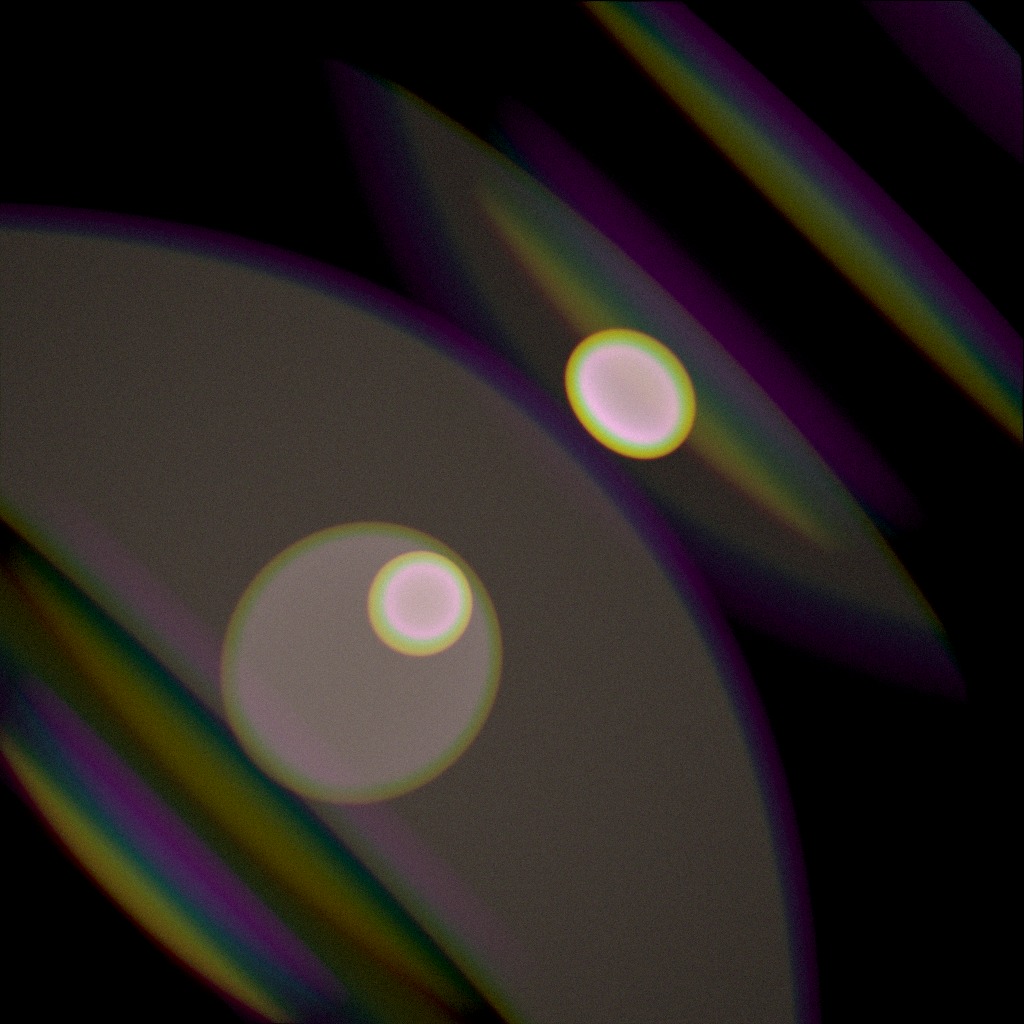}
            \end{minipage}\\
            & RT & Ours
        \end{tabular}
    \end{minipage}%
    \hfill
    \begin{minipage}{0.33\textwidth}
        \centering
        \includegraphics[width=\linewidth]{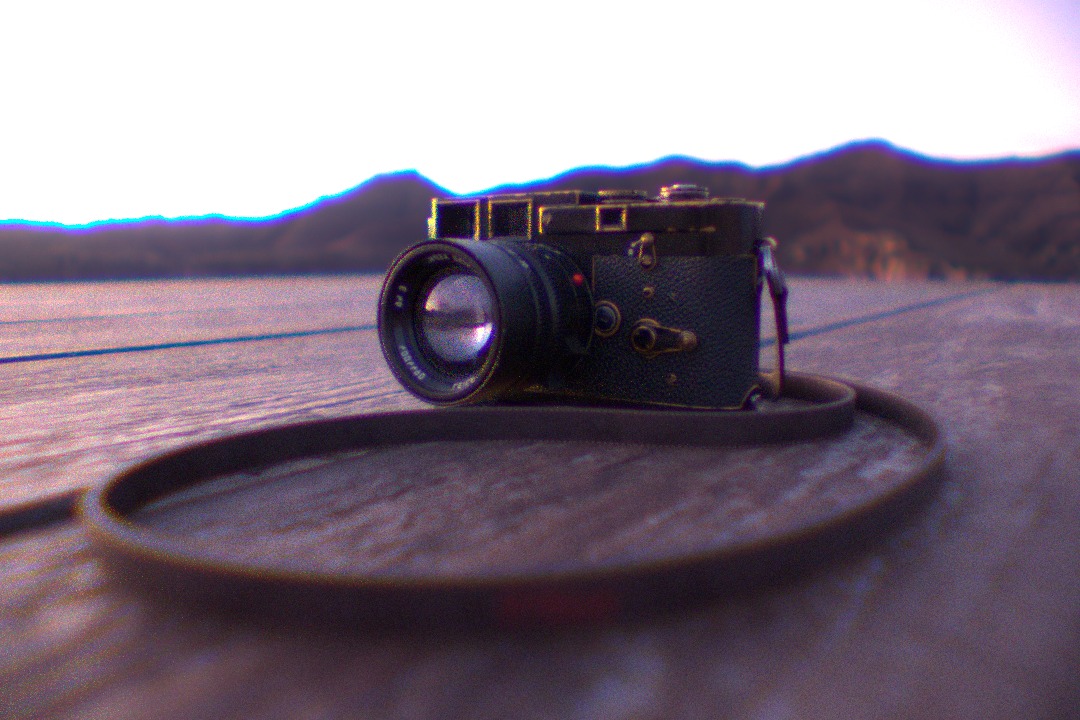} \\
        {\small 24mm Lens. Scene “Camera” \\(32768spp, Ours 74.4s/RT 1180.3s)}
    \end{minipage}
    \caption{Left: Overview of our forward pass pipeline for lens flare rendering. Middle: Comparison of lens flare results between ray tracing (RT) and our neural method (Ours) for 59\,mm (top) and 22\,mm (bottom) lenses. Right: Backward path tracing result for a camera-in-scene setup using a 24\,mm lens.}
    \label{fig:teaser}
\end{teaserfigure}

\maketitle

\section{Introduction}
Simulating light transport through lens systems has many applications in computer graphics, vision, and optics. 
Most existing approaches rely on ray tracing, which accurately models light propagation by tracing individual rays as they interact with complex lens surfaces. 
While general, ray tracing through a lens system adds non-negligible computation cost over simple models such as the pinhole camera or the thin lens, with costs scaling significantly as the number of lenses and surface complexity increases \cite{Steinert2011LensSimulation}. 
This additional cost has prevented widespread use of such realistic lens systems in place of those other simple models for realistic image synthesis. 
Prior work thus focused on developing a fast and accurate approximation of ray tracing through lens systems. 

Following the simplest linear model called \emph{ABCD} matrices, a majority of prior work focused an idea of approximating a map from input rays to output rays. 
This map can then replace the ray tracing process through a lens system given input rays. 
The ABCD matrices model this map as a linear transformation of an input ray (origin and direction) to an output ray, which is accurate only around the central axis of a given lens system. 
\citet{DBLP:journals/cgf/HullinHH12} proposed to use a polynomial to express this map for each lens and concatenate them with truncation to form a lens system, yielding an accurate approximation for origins and directions of output rays through a lens system. 
\citet{Hanika:2014:Lens} improved this polynomial model via nonlinear optimization~\cite{Levenberg:1944:MSC} and demonstrated how this fitted polynomial model can be used for rendering with depth of field effects. 
\citet{Zheng:2017:NeuroLens} used a neural network to further improve its accuracy for specific cases of refraction-only light paths.  
Those prior work have demonstrated the effectiveness of the core idea of modeling mapping from input rays to output rays, but come with various simplifications. 

Common to all those prior work, none of them provided a concrete and accurate model for encoding the Fresnel throughput which is quite important for realistic rendering of lens flares~\cite{DBLP:journals/tog/HullinESL11}.  
Wavelength-dependent transport through a lens system is quite important to capture its chromatic aberrations. 
While some prior work~\cite{Zheng:2017:NeuroLens,Hanika:2014:Lens} account for this wavelength-dependent nature of mapping by taking a wavelength as an input to the model, the wavelength-dependent nature of the Fresnel throughput, again, has not been modeled. 

Another challenge is that light paths going through a lens system can be occluded due to apertures and housing which introduces discontinuities in the mapping. 
The polynomial model~\cite{DBLP:journals/cgf/HullinHH12,Hanika:2014:Lens} approximates aperture occlusion by additional polynomials, whereas the domain subdivision technique used for the neural model~\cite{Zheng:2017:NeuroLens} was demonstrated only for refraction-only light paths (i.e., cannot render lens flares). 
No single model so far can accurately capture occlusion, Fresnel throughput, and wavelength-dependent transport, all at once to cover various lens effects such as bokeh effects, chromatic aberrations, and lens flares. 

We introduce a new model for light transport within a lens system, \emph{precomputed lens transport maps}, which can capture all those effects in a single model. 
Like the prior models~\cite{DBLP:journals/cgf/HullinHH12,Hanika:2014:Lens,Zheng:2017:NeuroLens}, we precompute and model the mapping between input rays and the resulting distribution of output rays emerging from a lens system. 
We use a neural network to model this mapping \emph{and} the Fresnel throughput for the first time to fully capture wavelength-dependent transport through a lens system. 
Recognizing the highly discontinuous nature, we propose to use a \emph{factorized} network, where one network is responsible for \emph{binary} classification of occlusion and the other network is responsible for representing a \emph{smooth} map. 
The previous polynomial model and the neural model are fundamentally designed to represent the latter smooth part in mind, and our model is the first to directly capture the discontinuity without any further approximations like modeling via polynomials or refraction-only paths, limiting their scopes of applications.   
Our precomputed transport map can instead be used in various applications, from realistic camera simulation to lens flare rendering, which previously needed different models for different applications. 
To summarize, our contributions are

\begin{itemize}

    \item \textbf{Accurate modeling of full light transport}: Unlike previous approaches, our method models both Fresnel refraction and reflection with wavelength-dependent effects throughout complex lens systems, enabling accurate simulation of lens flares and other high-order optical phenomena.
    
    \item \textbf{Consistent regressor accuracy}: Our regressor achieves lower and more uniform error across all valid light transport paths compared to existing approximation methods, providing reliable accuracy for both common and rare ray trajectories without artifacts through the use of $C^1$ continuous network.
    
    \item \textbf{Explicit classifier-based occlusion modeling}: We introduce a dedicated classifier that efficiently bounds the input space by filtering out invalid or occluded paths. This explicit occlusion modeling significantly improves both efficiency and generalization.

\end{itemize}

\section{Related Work}
\label{sec:related_work}
\paragraph{Ray tracing lens simulation}
While it is an approximation due to geometric optics, ray tracing has been used for simulating light transport through a lens system. 
\citet{DBLP:conf/siggraph/KolbMH95} proposed to replace the pinhole camera or the thin lens model by ray tracing through a lens system. 
Their results demonstrated visual significance of properly modeling light transport through a lens system. 
\citet{DBLP:journals/cgf/JooKLEL16} similarly demonstrated effects from aspherical lenses. 
In both cases, Fresnel reflections are not simulated which are important for accurate rendering of lens flares~\cite{DBLP:journals/tog/HullinESL11,lee2013practical}.
While more accurate simulation is possible~\cite{Steinert2011LensSimulation}, its computation cost generally scales with the number of lenses. 
The results of such simulation serves as training data for our model, and our aim is to replace ray tracing by a simpler evaluation of a neural network model. 

\paragraph{Lens system as mapping}
Besides ray tracing, one conventional approach to model light transport through a lens is based on a linear map between input rays and output rays under the paraxial assumptions. 
In this case, input ray origins and directions are multiplied by a matrix (called \emph{ABCD matrix}) to map it to a corresponding output rays. 
\citet{lee2013practical} used this linear model to simulate lens flares, though they noted its inaccuracy beyond certain angles toward a lens system. 
Polynomial optics \cite{DBLP:journals/cgf/HullinHH12} generalized this linear model to a polynomial model by performing a multivariate Taylor expansion of the analytical transport at the center axis of a given lens. 
A complete lens system can be formed by concatenation and truncation of each individual polynomial per lens. 
This polynomial representation enables a compact and accurate presentation of a mapping of transport through a lens system, which led to various follow up work such as sparse polynomial representations~\cite{schrade2016sparse} and a combination with domain subdivision~\cite{zheng2017adaptive}. 
\citet{Hanika:2014:Lens} further improved the precision of polynomial optics by fitting the polynomial directly to ground-truth ray tracing, making its application possible for depth of field rendering. 
The polynomial model has been used in practical applications for offline rendering for feature films~\cite{pekkarinen2019physically} and realtime rendering of lens flares~\cite{bodonyi2024real}. 
More recently, \citet{Zheng:2017:NeuroLens} showed that neural networks can improve accuracy when combined with domain subdivision~\cite{zheng2017adaptive}, but focused only on one-to-one mapping from incident to exitant ray, not including reflection (i.e., does not render lens flares). 
We conjecture that it is because reflected light paths often have different discontinuities than refraction-only paths due to occlusion, needing to have one domain subdivision per path type. 
Our model also uses a nonlinear mapping modeled by a neural network, but we proposed to use a \emph{factorized} representation where we have a product of the outputs of classifier and regressor networks, without any domain subdivision. 
This factorization allows us to capture both refraction and reflection with occlusion, enabling our model to support both lens flare and depth of field rendering under a single model. 

One missing factor is encoding of intensity changes through the Fresnel term at each lens surfaces which we called as \emph{Fresnel throughput} in this paper. 
Accurate estimation of the Fresnel throughput is crucial for getting the brightness right for image rendering as well as capturing intricate patterns of lens flare patters~\cite{DBLP:journals/tog/HullinESL11,lee2013practical}. 
While it is possible to compute this factor exactly~\cite{pekkarinen2019physically}, most reflected light can either ended up being occluded or have significantly low energy with the amount of computation scales with the number of lenses, thus wasting computation for (nearly) zero throughput. 
As such it is desirable to encode the Fresnel throughput as a precomputed model to avoid encoding such occluded or low energy paths in the mapping. 
\citet{DBLP:journals/cgf/HullinHH12} briefly noted how they can still fit polynomials, but we unfortunately could not find any further details. 
It is also unclear if polynomials are good fit for the Fresnel throughput and \citet{Hanika:2014:Lens} noted it as a limitation for their polynomial model. 
Our network encodes the Fresnel throughput within the same network for mapping, allowing us to efficiently and accurately represent the intensity change. 

\paragraph{Lens design}
Light transport simulation through a lens system plays an important role for prototyping lens systems. 
One common approach is to analyze the point spread function (PSF) for a given point in a scene that for a certain pattern on a sensor after going through a lens system~\cite{rossmann1969point}. 
The PSF alone, however, would not be able to model occlusion and occlusion and interreflections, thus ray tracing simulation is still viable. 
Most recently, \citet{teh2024aperture} demonstrated the use of differentiable ray tracing for automated optimization of a lens system.  
They noted existing approximations such as the polynomial model~\cite{DBLP:journals/cgf/HullinHH12} and the neural model~\cite{tseng2021differentiable} do not have enough accuracy for the purpose of their optimization. 
While our model does not aim for such lens optimization, as we demonstrate later, our model provides accurate approximation of ray tracing including Fresnel throughput for the first time, making it an attractive model for lens optimization as future work.

\section{Background}
A lens system can be described as a \emph{lens transport map $T$} that transforms an incident ray into a set of outgoing rays with associated intensity values. 
A single input ray will split into multiple output rays due to Fresnel reflections and refractions. 
Let $\bm{p}_{in},\bm{p}_{out}  \in \mathbb{R}^3$ denote the position on the input and output plane (before and after a lens system), $\bm{\omega}_{in},\bm{\omega}_{out} \in \mathbb{S}^2_+$ as the unit direction vector at input position and out position, $\lambda \in \mathbb{R}^+$ as the wavelength of light, and $I_{in}, I_{out} \in \mathbb{R}^+$ as the Fresnel throughput
\begin{equation}
    \mathcal{T}: (\bm{p}_{\text{in}}, \bm{\omega}_{\text{in}}, I_{in}, \lambda) \rightarrow \{(\bm{p}_{\text{out}}^{(k)}, \bm{\omega}_{\text{out}}^{(k)}, I_{out}^{(k)})\}_{k=1}^{n}
\end{equation}
where $n$ represents the number of output rays generated through multi-bounce interactions, and  $k$ is an index to a type of light path which will be defined later.

\begin{figure}[t]
    \centering
    \includegraphics[width=1\linewidth]{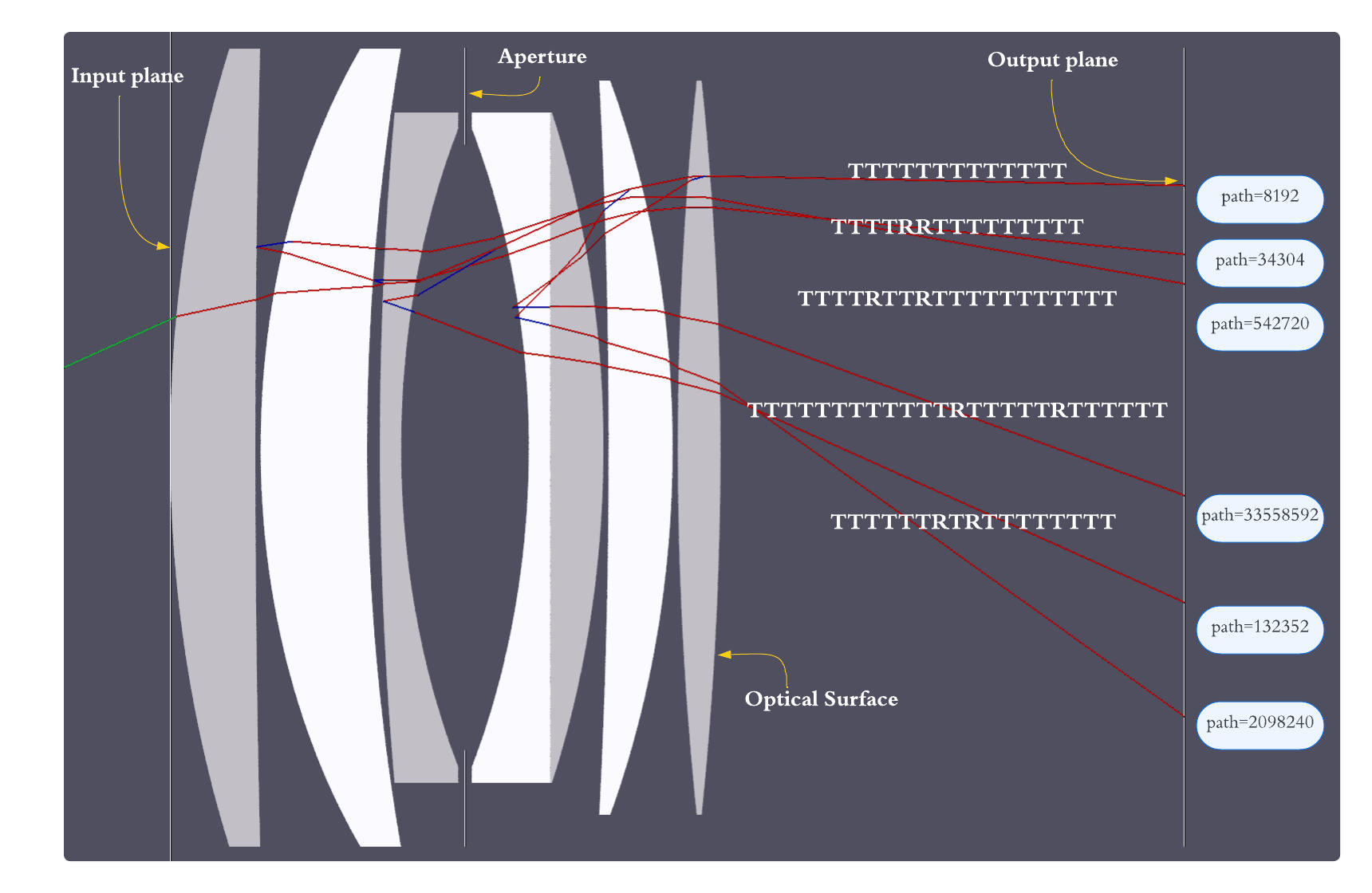}
    \caption{We decompose light transport paths in a lens system into many types of paths, each identified using a sequence of scattering decisions. Each sequence can be converted to a binary number that identifies the type. }
    \label{fig:path}
\end{figure}

\paragraph{Surface Interaction Model} 
Fig.~\ref{fig:path} illustrates the multi-valued nature of the lens transport mapping ($n \ge 1$) from involving reflections and refractions. 
The number of output rays varies with configurations of input rays because paths can intersect non-lens boundaries in the middle (e.g., housing or aperture) during ray tracing, making them absorbed and not reaching the other end of a lens system. 

During ray tracing, \textbf{ray state} represents a ray as:
\begin{align}
    y = 
\begin{bmatrix}
x & \omega &I&\lambda
\end{bmatrix}^{T},
\end{align}
where $ x \in \mathbb{R}^3$ is the ray origin, $\omega\in \mathbb{R}^3$ is the ray direction, and $I\in \mathbb{R}^+$ is the intensity, and $\lambda$ is the wavelength.
Let $\mathcal{L} \in \{T, R\}$ be the interaction type (transmission and reflection), and let the light path $\mathcal{P}$ be a sequence of interaction type 
\begin{align}
    \mathcal{P} = (\mathcal{L}_1, \mathcal{L}_2, \ldots, \mathcal{L}_K).
    \label{eq:light_path_def}
\end{align}
If there are $K$ interactions, then the ray must also intersect $K$ optical surfaces. We define each surface as $\sigma$, which can represent any type of surface, which includes optical surfaces, apertures, or the output plane (shown in Figure~\ref{fig:path}). If the ray continues to propagate
after interaction $\mathcal{L}_K$, it will eventually intersect the surface $\sigma_{K+1}$. We refer to the sequence of surfaces encountered by the ray as follows: 

\begin{align}
    (\sigma_1,\sigma_2, \ldots, \sigma_K, \sigma_{K+1}),
\end{align}
where the ray encounter optical surface $\sigma_1$ and then choose interaction $\mathcal{L}_1$, then it hits another optical surface $\sigma_2$ and choose interaction  $\mathcal{L}_2$, repeating this process until it hits the surface $\sigma_{K+1}$ which can any kind of surface.

Given a specific input ray $(x_{\text{in}}, \omega_{\text{in}}, \lambda)$ and a light path $\mathcal{P}$ consisting of $K$ interactions, we can perform ray tracing and the ray will hit the surface $\sigma_{K+1}$. At this point, we have a new ray whose position lies at the intersection with $\sigma_{K+1}$. This ray is considered one of the outputs of the lens transport mapping $\mathcal{T}$ if and only if $\sigma_{K+1}$ is the output plane. In this case, we also say that $(x_{\text{in}}, \omega_{\text{in}}, \lambda)$ is \emph{valid}.

\paragraph{Composite Operator Formulation}
We model ray tracing as a composite of three operators, positional operator $p_\sigma$ which determine the closest hit on the surface $\sigma$ to the ray origin, directional operator $d_{\mathcal{L},\sigma}$ which determine new direction after interaction $\mathcal{L}$, and the Fresnel function $f_{\mathcal{L},\sigma}(y_{in})$ for the Fresnel term. 
Suppose we know the ray $x_{in}$ will hit the surface $\sigma$, then we can define the path transport operator $S$ that will transform it to a new ray after interaction $\mathcal{L}$ and it has the form of
\begin{align}
    y_{out}=S_{\mathcal{L},\sigma}(y_{in}) = 
\begin{bmatrix}
p_{\sigma}(x_{in},\omega_{in},\lambda)\\
d_{\mathcal{L},\sigma}(x_{in},\omega_{in},\lambda) \\
f_{\mathcal{L},\sigma}(\omega_{in},\lambda, I_{in})
\end{bmatrix}.
\end{align} 

Given an initial ray state $y_{in}$ and a unique path $\mathcal{P} = (\mathcal{L}_1, \ldots, \mathcal{L}_K)$, the propagation through the system is determined by a sequence of path transport operator $\{S_{{\mathcal{L}_k,\sigma_k}}\}$. 
The composite operator for the full path is
\begin{align}
    \mathcal{T}^{\mathcal{P}} =  S_{{\mathcal{L}_k,\sigma_k}}  \circ \cdots \circ S_{{\mathcal{L}_2,\sigma_2}} \circ S_{{\mathcal{L}_1,\sigma_1}},
     \label{eq:composite_operator}
\end{align}
and the output state $y_{out}$ of that unique path $\mathcal{P}$  is given by
\begin{align}
y_{out}^\mathcal{P}=\begin{bmatrix}
x_{out}^\mathcal{P} &
\omega_{out}^\mathcal{P} &
I_{out}^\mathcal{P}&
\lambda
\end{bmatrix}^T = \mathcal{T}^{\mathcal{P}}(y_{in}).
\end{align}
Given this model, we consider forward light tracing and backward path tracing to model light paths going from either side from a given lens system. 
Due to its non-bijective nature, these two ways generally result in different mappings, thus one would need to compute a new map if a different direction is needed even for the same lens system. 

\paragraph{Forward Light Tracing}
Let the output plane be CMOS sized rectangle and let $\{\mathcal{P}_{\text{valid}}\}$ be a set that contains all the valid light path for the input ray $y_{in}$ that emits into the lens system, the final pixel intensity produced by this ray is computed by accumulating the energy transported along all valid light paths that reach the sensor:
\begin{align}
    I_{\text{cmos}}(y_{in})  = \sum_{\mathcal{P} \in \{\mathcal{P}_{\text{valid}}\}} I^{\mathcal{P}}_{out}(y_{in})\cdot h(x_{out}^{\mathcal{P}}(y_{in})) \cdot G,
    \label{eq:flare_integral_geo}
\end{align}
where $x_{out}^\mathcal{P}$ is the intersection point on the sensor and $h(\cdot)$ is a pixel filter. 
This formulation simply accumulates all rays that arrive at the pixel, weighted by their transported intensity and projected geometric factor $G$, which is a useful formulation for lens flare rendering.  

\paragraph{Backward Path Tracing}
Suppose the output plane is still the sensor plane and the input plane is the virtual plane in front of the lens system. 
The pixel intensity is computed by integrating over all directions within the hemisphere above the sensor point. 
For each sampled direction, a ray is traced through the lens system, considering only paths with full transmittance path $\mathcal{P}^{Transmit}$ (i.e., without internal reflections). 
Each valid path is then continued into the scene, and do general rendering. Formally, the pixel intensity is given by:
\begin{align}
    I_{\text{pixel}}(\lambda) = \int_{\Omega^+} L_{\text{scene}}(\mathcal{T}_{Transmit}^{-1}(x_{\text{sensor}},\omega,1,\lambda))  \cos\theta \, d\omega,
    \label{eq:backward_path_tracing}
\end{align}
where $\Omega^+$ is the upper hemisphere centered at $x_{\text{sensor}}$, $L_{\text{scene}}(\omega)$ is the radiance arriving from the scene along direction $\omega$, $T(x_{\text{sensor}}, \omega)$ is an indicator function that is $1$ if the traced path through the lens is fully transmitted and $0$ otherwise, and $\theta$ is the angle between $\omega$ and the sensor normal.  
This model is useful for rendering from a camera through a lens system (e.g., depth-of-field effect). 

\section{Precomputed Lens Transport Maps}
The existing approaches to lens system rendering suffer from a few restrictions; direct ray tracing can be slow for complex lens systems, polynomial models lack sufficient accuracy near lens edges~\cite{DBLP:journals/cgf/HullinHH12} or lack proper encoding of the Fresnel throughput, and neural model~\cite{Zheng:2017:NDC} requires complicated domain subdivision and network ensembles to account for discontinuities in the mappings. 
To overcome these issues, we introduce a new model via a \emph{factorized} neural network model, which we call as a \textbf{classifier-regressor structure}.
The classifier determines whether a ray produces a valid output, and the regressor predicts the exact output values. 
\begin{lstlisting}[float,language={python},label={list:nl-pf},mathescape=true,caption={Lens System Rendering Pipeline}]

def rendering pipeline():
    # Backward Path Tracing
    for each pixel:
        $p_{in}$ = sample on Backward Input Plane
        $\omega_{in}$ = connect pixel with sampled position
        
        is_blocked = Classifier($p_{in}, \omega_{in}, \lambda$)
        if is_blocked: continue
        
        ($p_{out}, \omega_{out}, I_{out}$)=Regressor($p_{in}, \omega_{in}, \lambda$)
        backward_color[pixel] = rayTracer($p_{out}, \omega_{out}$) \
        * $I_out$ * $dot(\omega_{out}, cmos_normal)$
    
    # Forward Light Tracing
    pointLight = locate brightest lightsource in scene

    for pathId in path combinations:
        for certain num of samples:
            $p_{in}$ = sample on Forward Input Plane
            $\omega_{in}$ = connect pixel with sampled position
            
            is_blocked = Classifier($p_{in}, \omega_{in}, \lambda$)
            if is_blocked: continue
            
            ($p_{out}, \omega_{out}, I_{out}$)=Regressor($p_{in}, \omega_{in}, \lambda$)
            color[pathId][$p_{out}$] +=   $I_{out}$ * $dot(\omega_{out}, n_{cmos})$

        color[pathId] /= samples

    finalImg = backwardColor + foreach forward_color
        
\end{lstlisting}

\subsection{Geometric Simplification} 
\label{sec:geo}
We assume that every lens in our optical system exhibits \emph{circular symmetry} (inherently implies \emph{reflection symmetries}), as this is the most common configuration in practical optical design. 
Furthermore, we assume that all optical axes of the lenses are aligned along a common straight line. 
These assumptions not only reflect typical real-world lens assemblies, but also simplify the mathematical modeling of light propagation and allow the lens transport mapping to inherit rotational and reflection symmetry about the optical axis. 
Note that its assumption is different from spherical lenses, and our model can still support aspherical lenses as long as they are circularly symmetric. 

Under this simplification, we can let $\mathcal{R}$ be either the any rotational transformation around the optical axis or any reflectional transformation respect the the plane that contain the the optical axis, the lens transport mapping has the following property:
$$
 \mathcal{T} (\bm{p}_{\text{in}}, \bm{\omega}_{\text{in}}, \lambda) =\{(\bm{p}_{\text{out}}^{(k)}, \bm{\omega}_{\text{out}}^{(k)}, I^{(k)})\}_{k=1}^{n}
$$
\begin{align}
     \mathcal{T} (\mathcal{R} (\bm{p}_{\text{in}}), \mathcal{R} (\bm{\omega}_{\text{in}}), \lambda) =
 \{(\mathcal{R}(\bm{p}_{\text{out}}^{(k)}),\mathcal{R} ( \bm{\omega}_{\text{out}}^{(k)}), I^{(k)})\}_{k=1}^{n} .
\end{align}
Because of these two symmetries, we can restrict the domain of our input lower dimensional spaces. For $p_{in}$, we restrict its domain from $\mathbb{R}^2$  to $\mathbb{R}_{\ge0}$ due to circular symmetry. For $\omega_{in}$, we restrict the domain from hemisphere space $S_{+}^2:\{\begin{bmatrix}x,y,z \end{bmatrix}^T \in \mathbb{R}^3: x^2 +y^2+z^2 = 1,y \ge 0\}$ to quarter-sphere space $S_{1/4}^2:\{\begin{bmatrix}x,y,z \end{bmatrix}^T\in \mathbb{R}^3: x^2 +y^2+z^2 = 1,x \ge 0,y\ge 0\}$ due to reflection symmetry. These two restrictions force the symmetry property of our model and further improves image quality.

\subsection{Neural Ray Mapping} 

\paragraph{Path Decomposition}
Given the variable number of physically valid light paths that a ray may follow through the lens system, we decompose the global lens transport mapping $\mathcal{T}$ into a set of path-specific mappings $\mathcal{T}^{\mathcal{P}}$, where each $\mathcal{P}$ denotes a unique sequence of surface interactions (refractions or reflections) as formalized in Equation~\eqref{eq:light_path_def}. This path decomposition is critical, as a given input ray may yield zero, one, or multiple valid outputs depending on geometry and occlusion conditions. 
This decomposition itself is the same as the one in the polynomial model~\cite{DBLP:journals/cgf/HullinHH12}.

Without this decomposition, one might naively consider training a single, large neural network that attempts to learn the mapping from inputs to all possible light transport paths simultaneously. 
One major issue is that such a network must output a variable number of predictions for each input, corresponding to every feasible path combination. 
It means constructing a multi-headed output, where each head represents a specific path configuration, and, during inference, the appropriate output head is selected based on the combination of active paths or their probabilities. 
As the number of possible paths grows, this naive approach leads to a dramatic increase in model size and complexity. 
Additionally, the highly discontinuous nature of the output space---with different paths becoming valid or invalid depending on the input---makes training such a network difficult and often unstable.

To address these challenges, we combine path decomposition with small networks: instead of a monolithic model, our pipeline processes each type individually. Each unique type is regressed separately, which avoids the combinatorial explosion in output space and simplifies the learning task for each model. Empirically, under typical absorption assumptions~\cite{DBLP:journals/cgf/HullinHH12,lee2013practical,Hanika:2014:Lens}, we observe that contributions from higher-order paths (with more than two bounces) are negligible, so we restrict our attention to the most relevant unique paths. If greater accuracy is needed, our method can be naturally extended to include additional higher-order paths. As illustrated in Fig.~\ref{fig:teaser}, all retained paths are processed independently, and their contributions are subsequently aggregated to produce the final rendered image.

\paragraph{Neural Ray Masking} 
\begin{figure}
    \centering
    \includegraphics[width=1\linewidth]{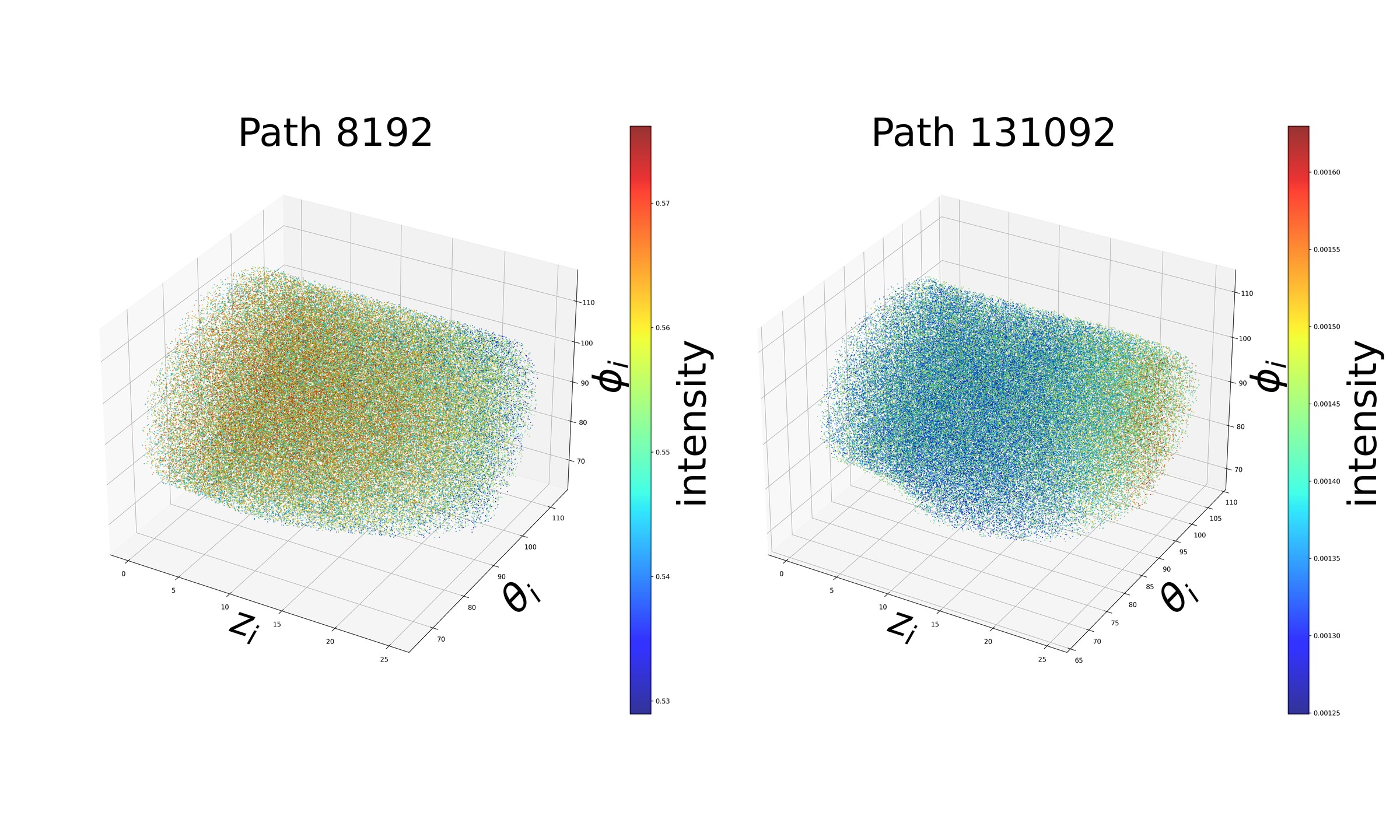}
    \caption{We plot the intensity for two type of path in the 3D input space. For each kind of path, valid intensity samples form a bounded manifold.}
    \label{fig:3DVis}
\end{figure}
Upon decomposing the lens transport mapping into path-specific mappings, we observe that not all inputs yield valid outputs for every path. For instance, rays may be blocked by the lens barrel or aperture, resulting in infeasible or undefined outputs. During neural network training, it is crucial that the network only fits the mapping within the valid region of the input space. 
The polynomial model~\cite{DBLP:journals/cgf/HullinHH12} handles such occlusion by introducing additional polynomials or explicit occlusion~\cite{pekkarinen2019physically} between a composition of mappings through each lenses. 
We propose to model occlusion through an \emph{entire} lens system, which naturally fits our neural approach. 

We find that the valid domain for each path forms a distinct, bounded manifold in the input space, as visualized in Figure~\ref{fig:3DVis}. To ensure the regressor operates exclusively on this feasible set, we introduce a binary mask using an MLP classifier: the classifier outputs $1$ if the input lies within the valid region for a given path, and $0$ otherwise. Let $x$ be the input, $f(x)$ the regressor, $g(x)$ the classifier, and $\{y\}$ a set of outputs from our model, we have:
\begin{align}
    \{y\}=
    \begin{cases}
    f(x), \text{ if } g(x) = 1\\
    \emptyset, \text{ otherwise }
    \end{cases}
    \tag*{,}
\end{align}
where $\emptyset$ is the empty set which mean the model won't output anything for this input $x$.
Formally, let $\Omega_{\bm{\mathcal{P}}}$ denote the set of all inputs for which path $\mathcal{P}$ does not encounter any edge discontinuity. The classifier thus partitions the input space into feasible ($\Omega_{\bm{\mathcal{P}}}$) and infeasible regions, ensuring that the neural network models the path-specific mapping only where it is physically valid.

\paragraph*{$C^1$ Continuous Network}
After path decomposition and neural masking have excluded all discontinuities, regression is confined to the valid region of the input space, where each input ray follows a unique, physically plausible path without abrupt state changes. In this region, it is important to characterize the mathematical properties of the light transport operator.
In the composite lens transport mapping $\mathcal{T}^{\mathcal{P}}$, as defined in Equation~\ref{eq:composite_operator}, each path is constructed from a sequence of path transport operators $S_{{\mathcal{L}_k,\sigma_k}}$ that are infinitely differentiable within $\Omega_{\bm{\mathcal{P}}}$. Under the assumption that sharp lens corners are blocked by the aperture, $\mathcal{T}^{\mathcal{P}}$ is at least $C^1$ (continuously differentiable) within the valid region. Since the path transport operators are continuous and the lens transport mapping $\mathcal{T}$ is a composition of these operators, the overall mapping $\mathcal{T}^{\mathcal{P}}$ retains continuity, as the composition of continuous functions remains continuous.
\begin{figure}
    \centering
    \includegraphics[width=1.0\linewidth]{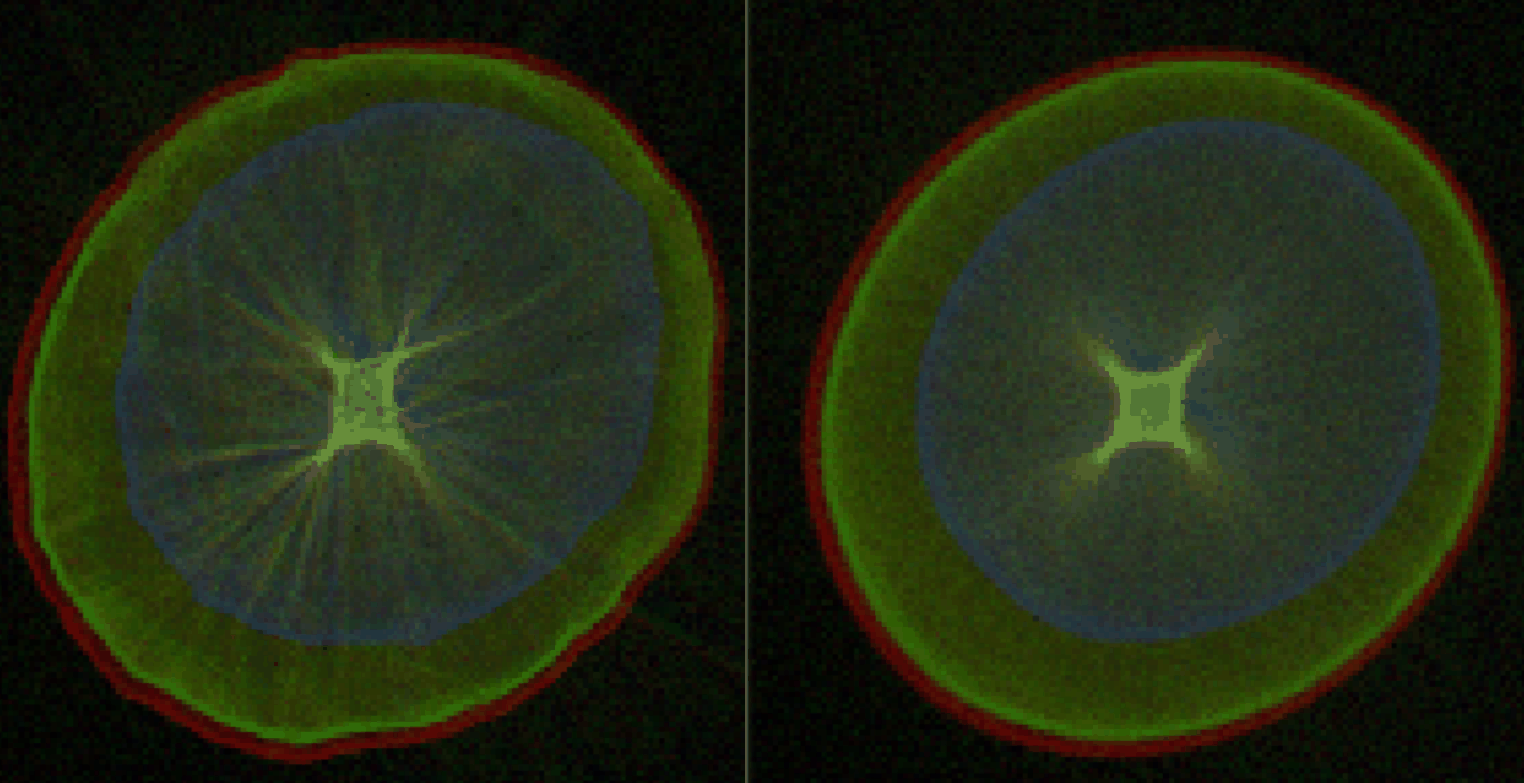}
    \caption{We compared the lens flares produced by a ReLU MLP and tanh MLP. Notice that there are many strip-like artifact in the left image,  and that the boundary of the left pattern resembles piecewise segments, due the piecewise linear nature of ReLU MLPs. Tanh MLPs on the other hand have continuous derivatives, leading to a much smoother pattern.}
    \label{fig:tanh}
\end{figure}
Given this $C^1$ continuity of the underlying mapping, it is necessary for the regression network to preserve the same level of smoothness. Networks based on ReLU activations, being piecewise linear, introduce gradient discontinuities that can produce visible artifacts in the rendered output. In contrast, the Tanh activation function is $C^1$ continuous, enabling the network to more faithfully approximate the underlying physics. As demonstrated in Figure~\ref{fig:tanh} . Tanh-based networks consistently yield smooth predictions, whereas ReLU-based networks introduce noticeable striping and piecewise linear boundaries. A more rigorous analysis of the continuity requirements for the regression network is deferred to future work.

\section{Results}
\label{sec:results}
\subsection{Implementation}
Our implementation is organized into three main stages: data collection via ray tracing, MLP training, and efficient inference integration for rendering.

\subsubsection*{Data Collection}
We implemented a custom lens ray simulation framework capable of handling arbitrary lens system. The lens configurations are provided as JSON files exported from the Open Optical Designer \cite{openopticaldesigner}. 
To train the regressor, we uniformly sample $(p_{in},\omega_{in})$ within the valid region, while the wavelength $\lambda$ is importance-sampled according to the CIE XYZ color matching function.
For each type of path, only a small fraction of the input space produces valid rays. To accelerate data generation, we employ a Markov Chain Monte Carlo sampler with a binary visibility target function \cite{hachisuka_robust_2011}.
Each type of path yields approximately 81 million valid samples, that occupy about 4.27~GB of storage. 
For the classifier, we generate an equal number of valid and invalid rays to ensure a balanced dataset.
The total storage required for the classifier data per path is approximately 1.85~GB. 

\subsubsection*{MLP Training}
We implemented all networks as compact MLPs, employing \verb|tanh| activations in the hidden layers and no activation at the output, with each layer consisting of 32 neurons. 
The regressor uses five hidden layers to improve prediction accuracy, with mean squared error (MSE) loss for position and intensity, and cosine similarity for direction. In contrast, the classifier, being a simpler task, uses only two hidden layers and is optimized using binary cross-entropy loss.
All models are trained with an initial learning rate of $10^{-4}$, which decays exponentially by a factor of 0.95 every 10,000 batches.
Each unique path regressor for light tracing is trained for 40 epochs with a batch size of 8192. For the full transmittance regressor, which requires higher precision as the ray propagates through the scene, training is conducted in two phases: an initial 200 epochs with a larger batch size of 32,768, followed by a 50-epoch fine-tuning stage with a smaller batch size (8192) and a reduced starting learning rate of $10^{-6}$.

\subsubsection*{Inference}
\begin{figure}
    \centering
    \includegraphics[width=1.0\linewidth]{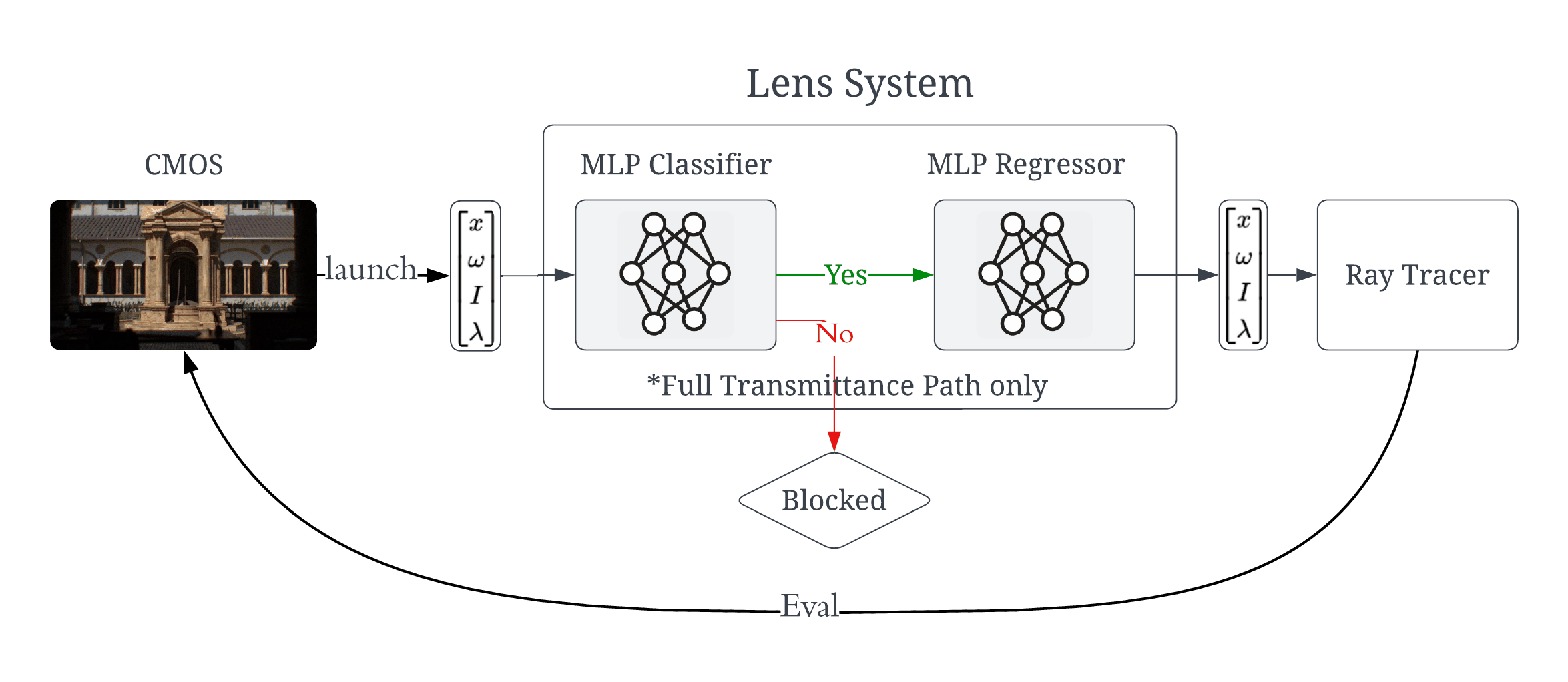}
    \caption{Overview of the backward path tracing pipeline used in our system. Rays are traced from the sensor through the lens elements to the scene, allowing simulation of optical effects such as aberrations and depth of field.}
    \label{fig:Path-Tracing-Pipeline}
\end{figure}

We integrated the brute force ray tracer and our method into the LuisaRender~\cite{Zheng2022LuisaRender} framework to enable GPU acceleration. 
The over pipeline is illustrated in Figure~\ref{fig:teaser} and Figure~\ref{fig:Path-Tracing-Pipeline}.
To efficiently query the network in GPU rendering, we fuse the MLP directly inside compute kernels. 
We also approximate tanh activations with a rational function to reduce computational cost. 

\subsection{Lens Flare Rendering}
\begin{figure*}[htbp]
    \centering
    \setlength{\tabcolsep}{1pt} 
    \begin{tabular}{c@{\hskip 3pt}ccc}
        & Poly & Ours & Reference \\
        \rotatebox[origin=c]{90}{\footnotesize 59mm} &
        \begin{minipage}{0.3\linewidth}
            \centering
            \includegraphics[width=0.85\linewidth]{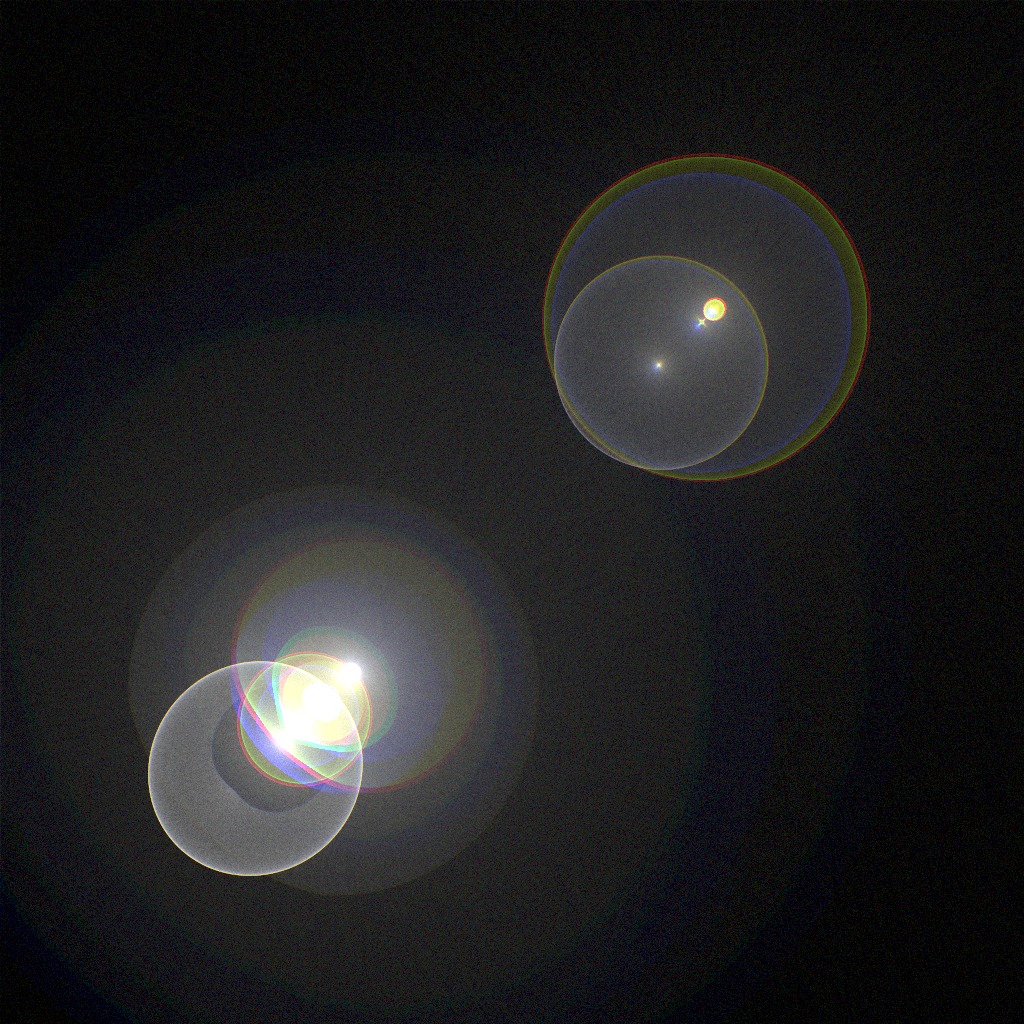}
        \end{minipage} &
        \begin{minipage}{0.3\linewidth}
            \centering
            \includegraphics[width=0.85\linewidth]{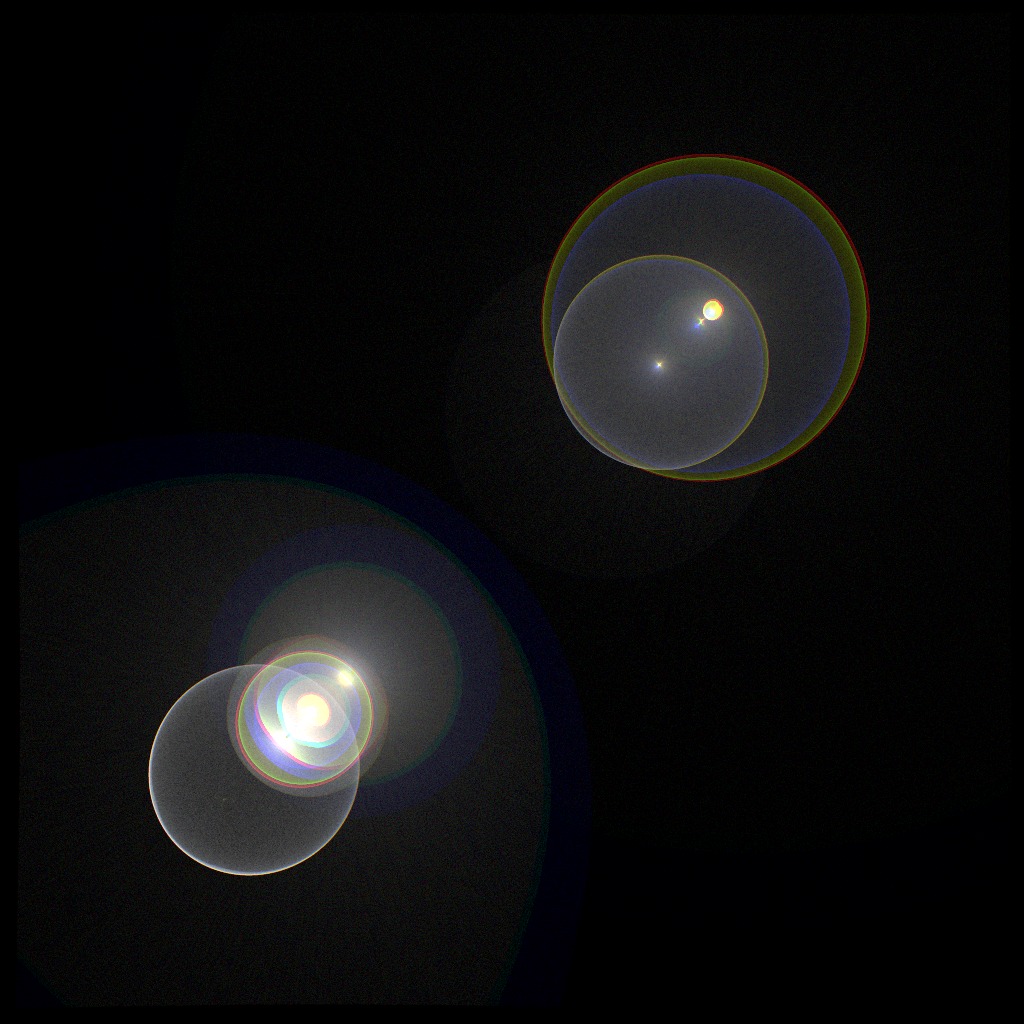}
        \end{minipage} &
        \begin{minipage}{0.3\linewidth}
            \centering
            \includegraphics[width=0.85\linewidth]{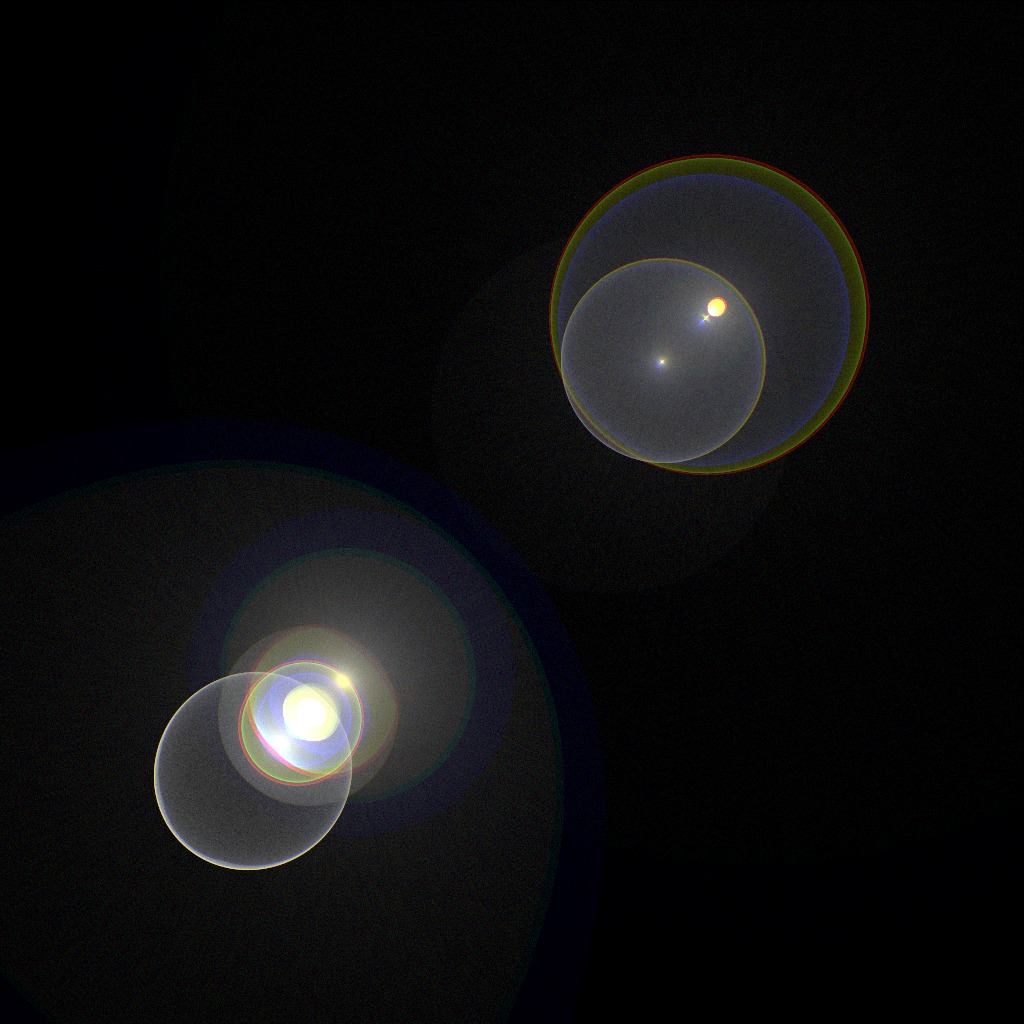}
        \end{minipage} \\
        MAPE& 0.148 & 0.047& -- \\
        \rotatebox[origin=c]{90}{\footnotesize 22mm} &
        \begin{minipage}{0.3\linewidth}
            \centering
            \includegraphics[width=0.85\linewidth]{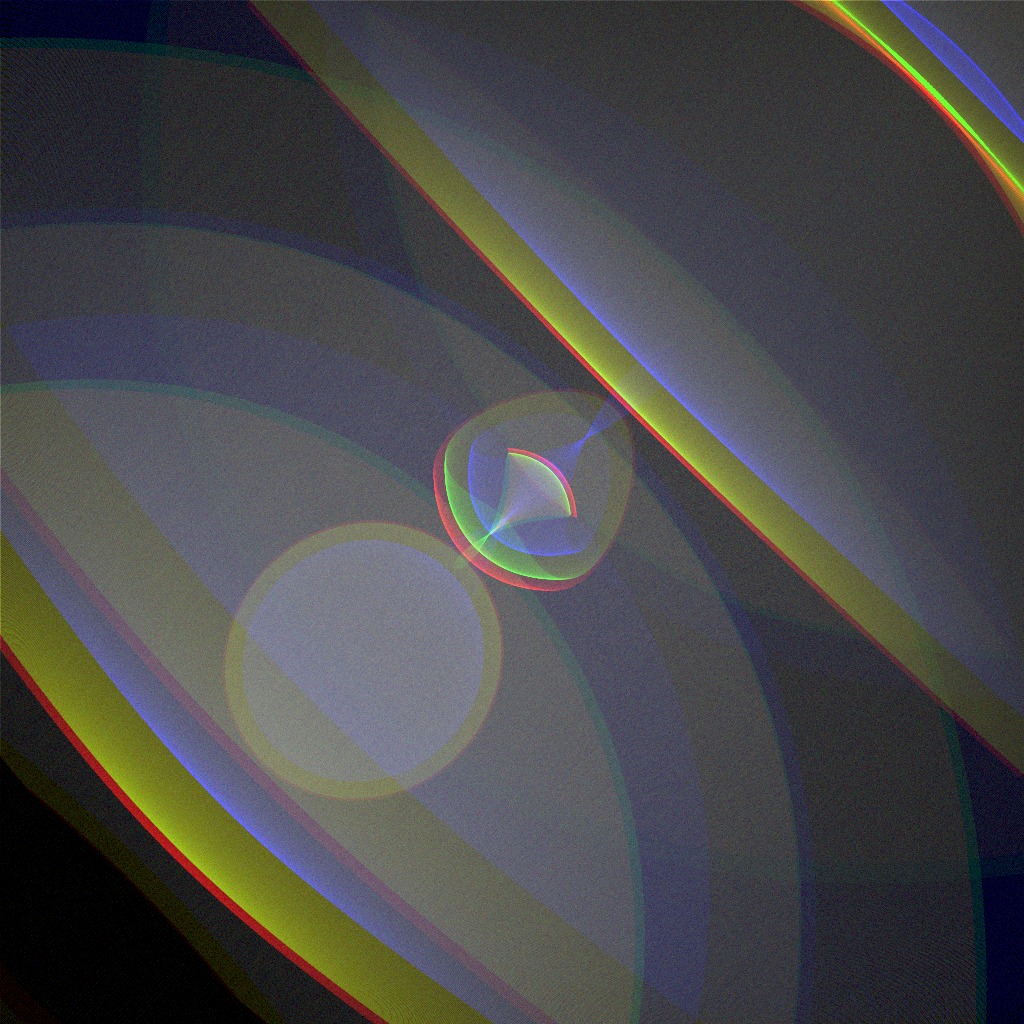}
        \end{minipage} &
        \begin{minipage}{0.3\linewidth}
            \centering
            \includegraphics[width=0.85\linewidth]{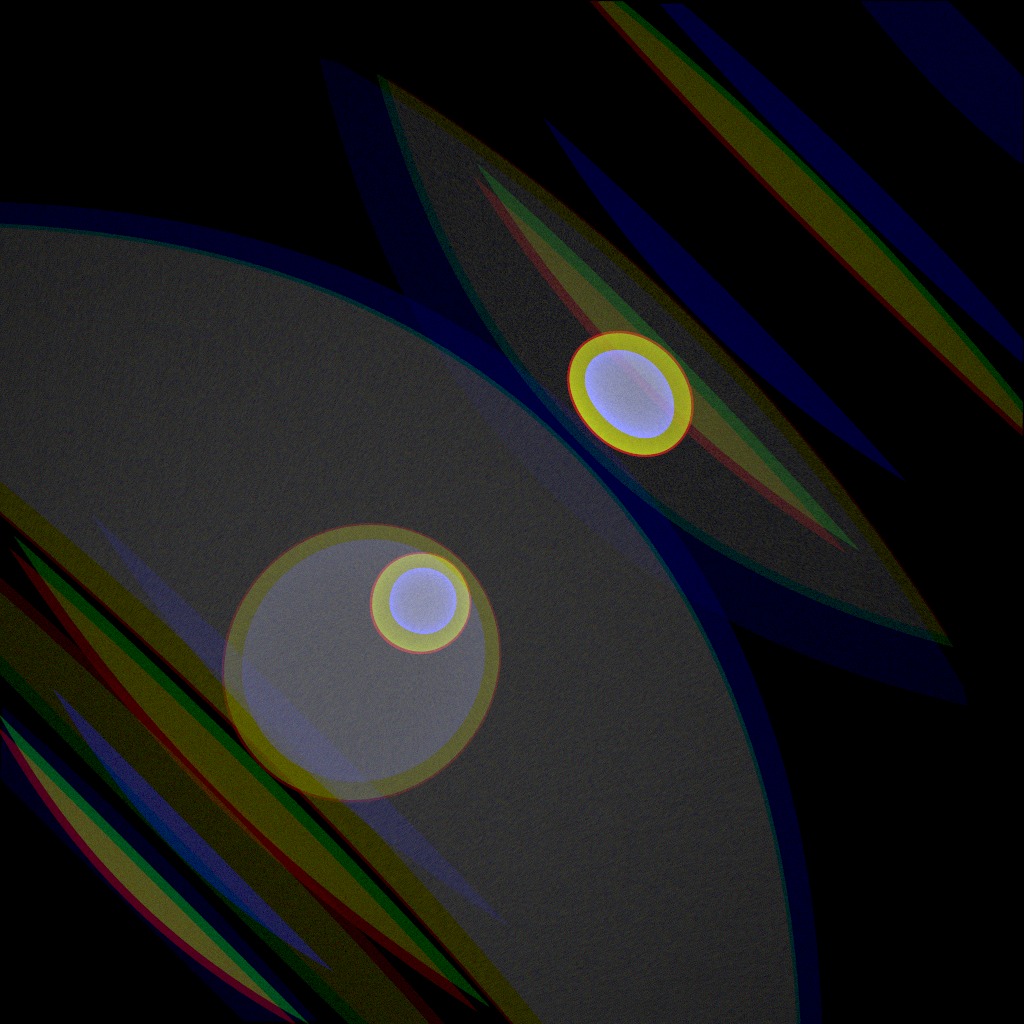}
        \end{minipage} &
        \begin{minipage}{0.3\linewidth}
            \centering
            \includegraphics[width=0.85\linewidth]{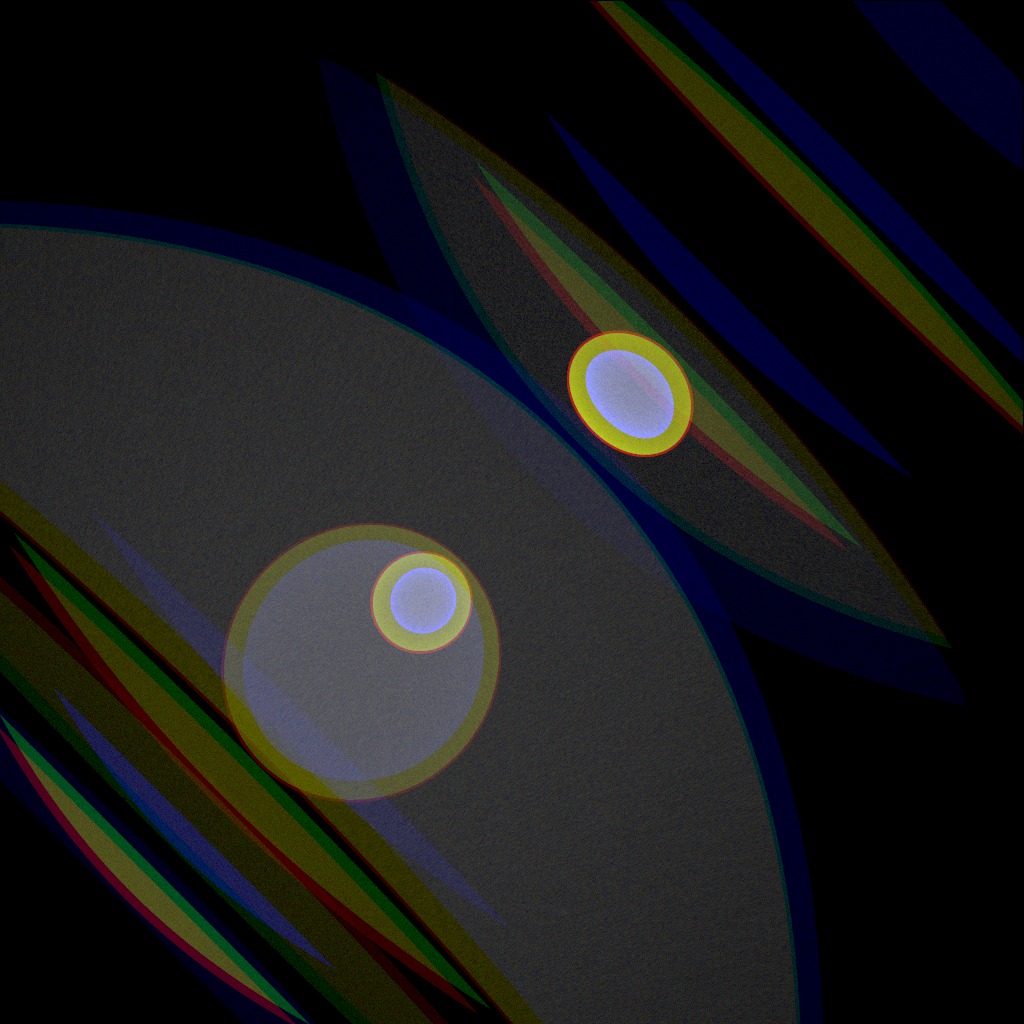}
        \end{minipage} \\
        MAPE& 0.694 & 0.032 & -- \\
        \rotatebox[origin=c]{90}{\footnotesize 59mm (path 131092)} &
        \begin{minipage}{0.3\linewidth}
            \centering
            \includegraphics[width=0.85\linewidth]{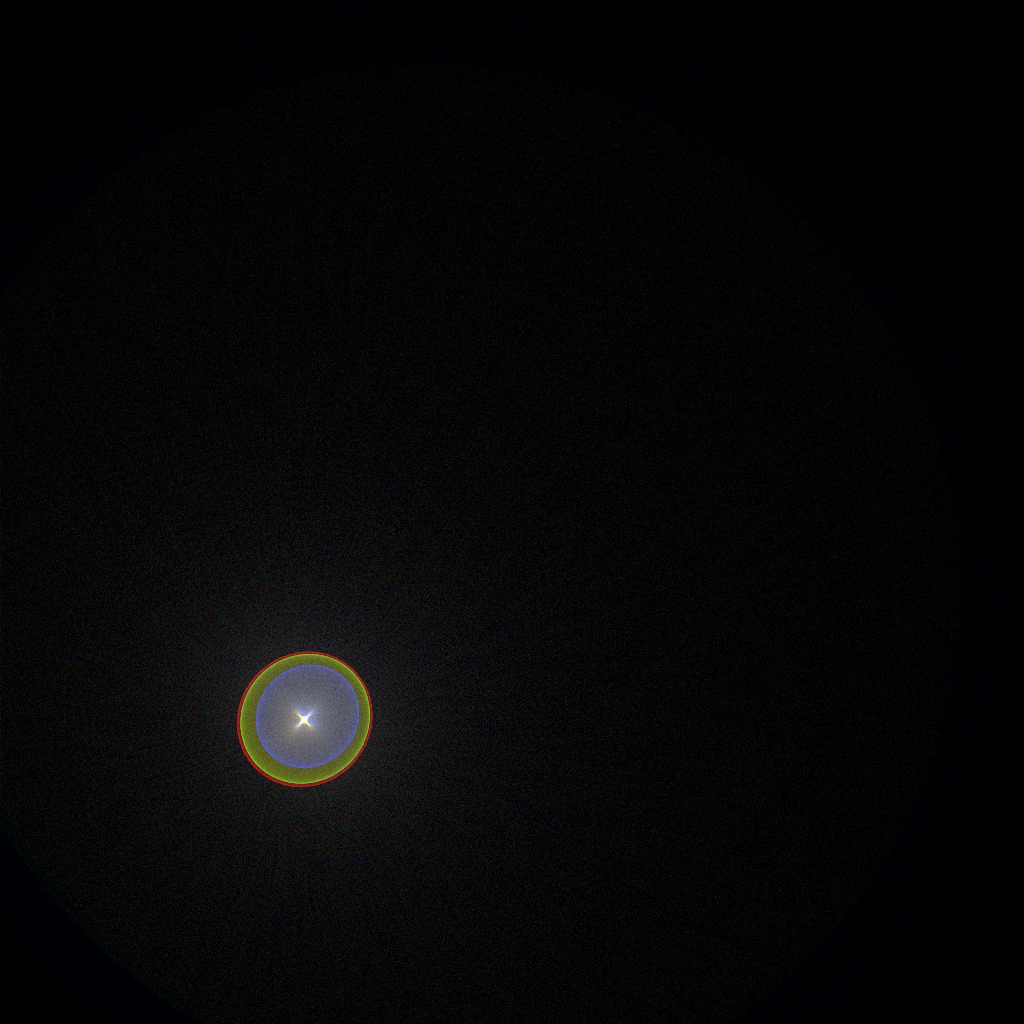}
        \end{minipage} &
        \begin{minipage}{0.3\linewidth}
            \centering
            \includegraphics[width=0.85\linewidth]{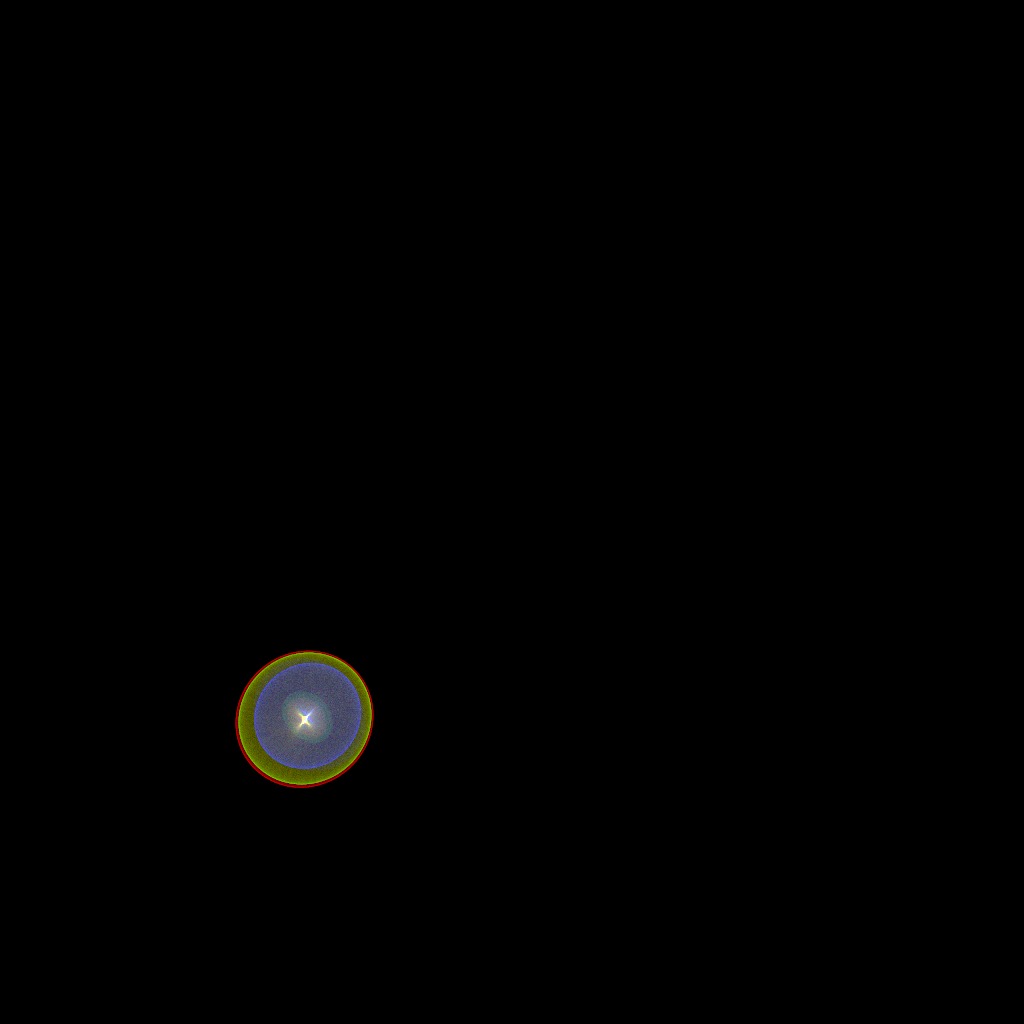}
        \end{minipage} &
        \begin{minipage}{0.3\linewidth}
            \centering
            \includegraphics[width=0.85\linewidth]{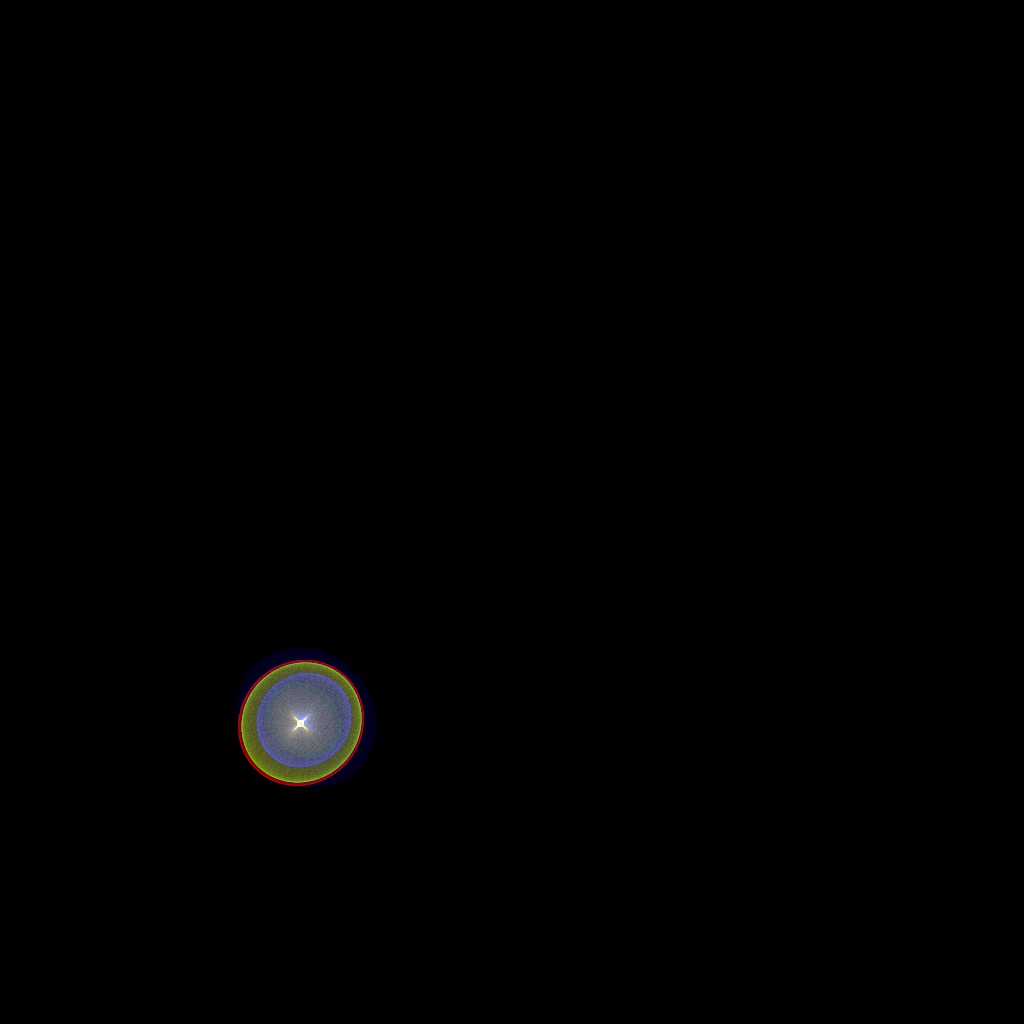}
        \end{minipage} \\
        MAPE & 0.027 & 0.007 & -- \\
        \rotatebox[origin=c]{90}{\footnotesize 22mm (path 65616)} &
        \begin{minipage}{0.3\linewidth}
            \centering
            \includegraphics[width=0.85\linewidth]{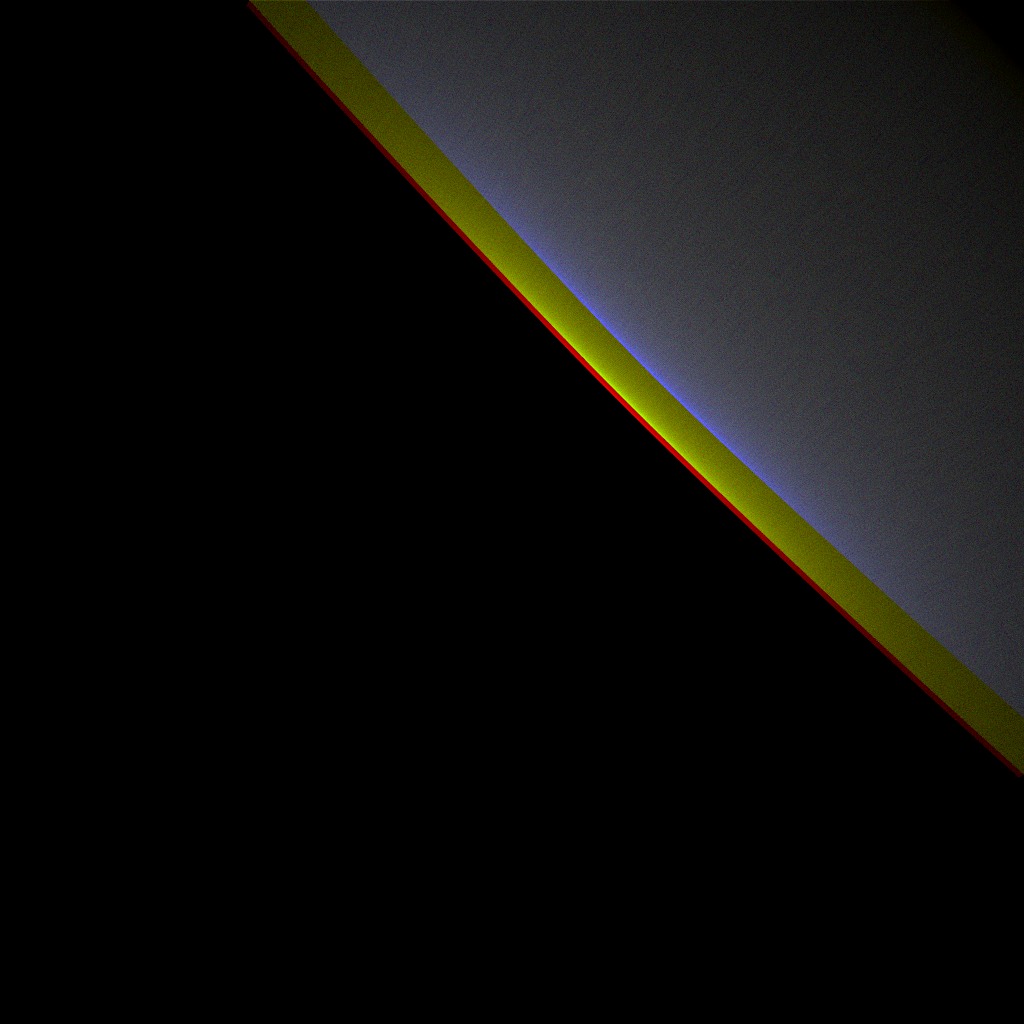}
        \end{minipage} &
        \begin{minipage}{0.3\linewidth}
            \centering
            \includegraphics[width=0.85\linewidth]{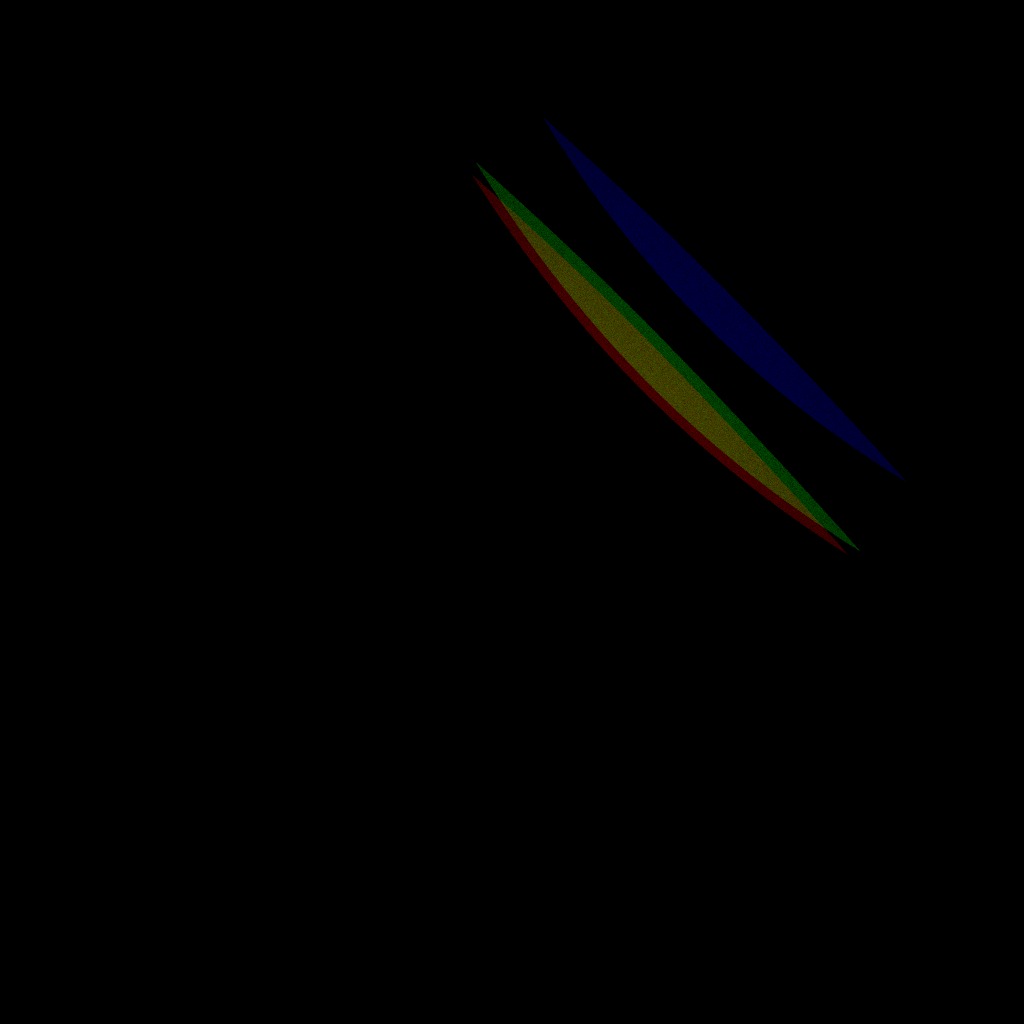}
        \end{minipage} &
        \begin{minipage}{0.3\linewidth}
            \centering
            \includegraphics[width=0.85\linewidth]{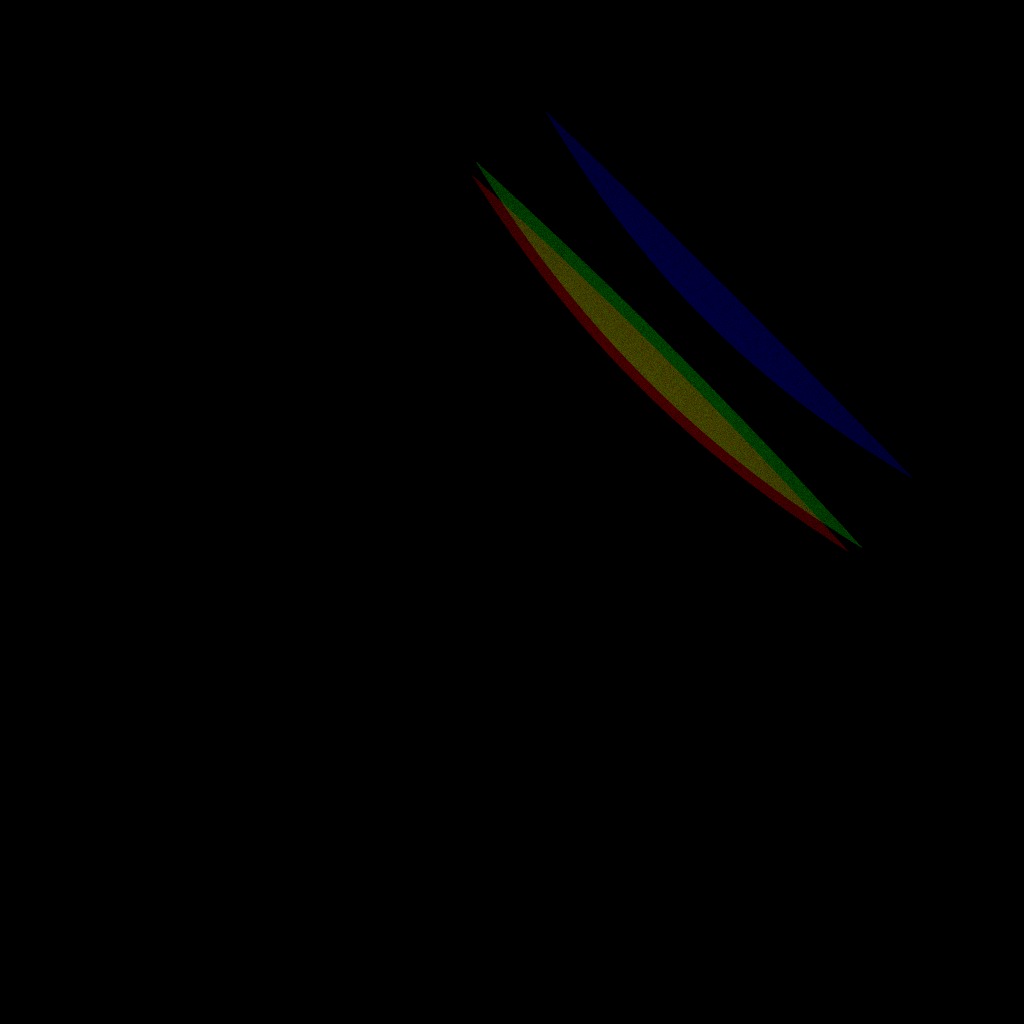}
        \end{minipage} \\
        MAPE & 0.147 & 0.002 & -- \\
    \end{tabular}
    \caption{Qualitative comparison of lens flare appearance for 59mm and 22mm focal lengths using different methods. The first two rows show results with all light paths combined for each focal length. The last two rows show results for a selected representative path: 59mm (path 131092, a case where the polynomial baseline succeeds) and 22mm (path 65616, a case where the polynomial baseline fails). Columns correspond to Poly (polynomial baseline), Ours (neural transport), and Reference (full simulation).}
    \label{fig:light_tracing}
\end{figure*}

We compared our method with Taylor polynomial optics \cite{DBLP:journals/cgf/HullinHH12} and the reference ray tracer in Figure~\ref{fig:light_tracing}.
For each image, we trace one million rays for each RGB channel and disable Fresnel throughput computation as they are not handled by polynomial approximations. 
Note that our method is capable of handling both continuous wavelength inputs and Fresnel throughput, as demonstrated in Figure~\ref{fig:teaser}.
Although polynomials approximate the lens transport well when paths are short, they struggle to fit longer paths, as shown in the last two rows of Figure~\ref{fig:light_tracing}. 
This inaccuracy for longer paths is caused by that the polynomial optics compute Taylor polynomials for each lens independently and concatenate them to form a lens system.  
As the number of lens involved in the light path length increases, the error in the approximation accumulates, eventually leading to inaccurate results when the input is far away from the Taylor expansion center. 
One can also observe that the polynomial results are much brighter than the ground truth as it does not model occlusion by the lens (barrel) housing in its original form. 
Many rays that should be occluded by the barrel are instead interacting with the lens and eventually come out of the lens system as valid rays and contribute the final image, resulting a brighter image. 
Such discontinuities cannot be handled efficiently by polynomial models, as they require fitting and evaluating an additional polynomial each time an occlusion test is performed~\cite{pekkarinen2019physically}.
We can also observe that lens flares of the 22mm lens rendered by polynomial model is less accurate compared to that of the 59mm lens since the 22mm lens has a wider FOV, resulting in a wider distribution of rays with more distorted rays on the edges. 
Our method is free from such issues and is robust under challenging lighting setups as demonstrated in the middle column of Figure~\ref{fig:light_tracing}.

\subsection{Depth of Field Rendering}
Our approach employs a single classifier--regressor network (the full transmittance pass) as the camera integrator, implemented as a plugin within the Luisa Renderer. As shown in Figure~\ref{fig:path-tracing}, we compare our neural lens transport with ground truth ray-traced lens systems across two scenes and three different lens designs. Quantitative results demonstrate that our method achieves very low mean absolute percentage error (MAPE), with all values below 0.15 and in many cases below 0.05, indicating that the neural approach closely matches the accuracy.

\subsubsection*{Manual Focusing}
\begin{figure}
    \centering
    \begin{tabular}{ccc}
        \includegraphics[width=0.3\linewidth]{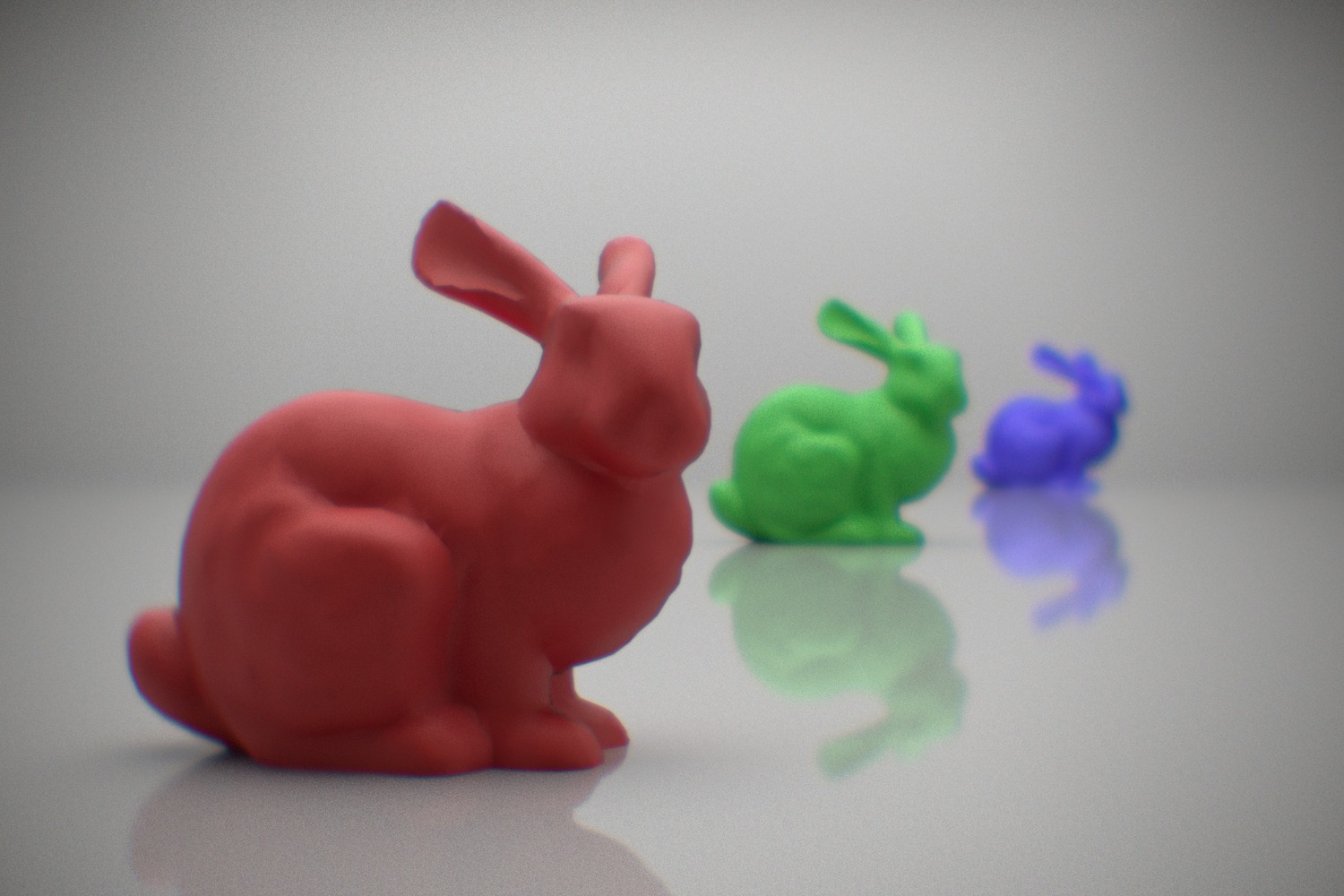} &
        \includegraphics[width=0.3\linewidth]{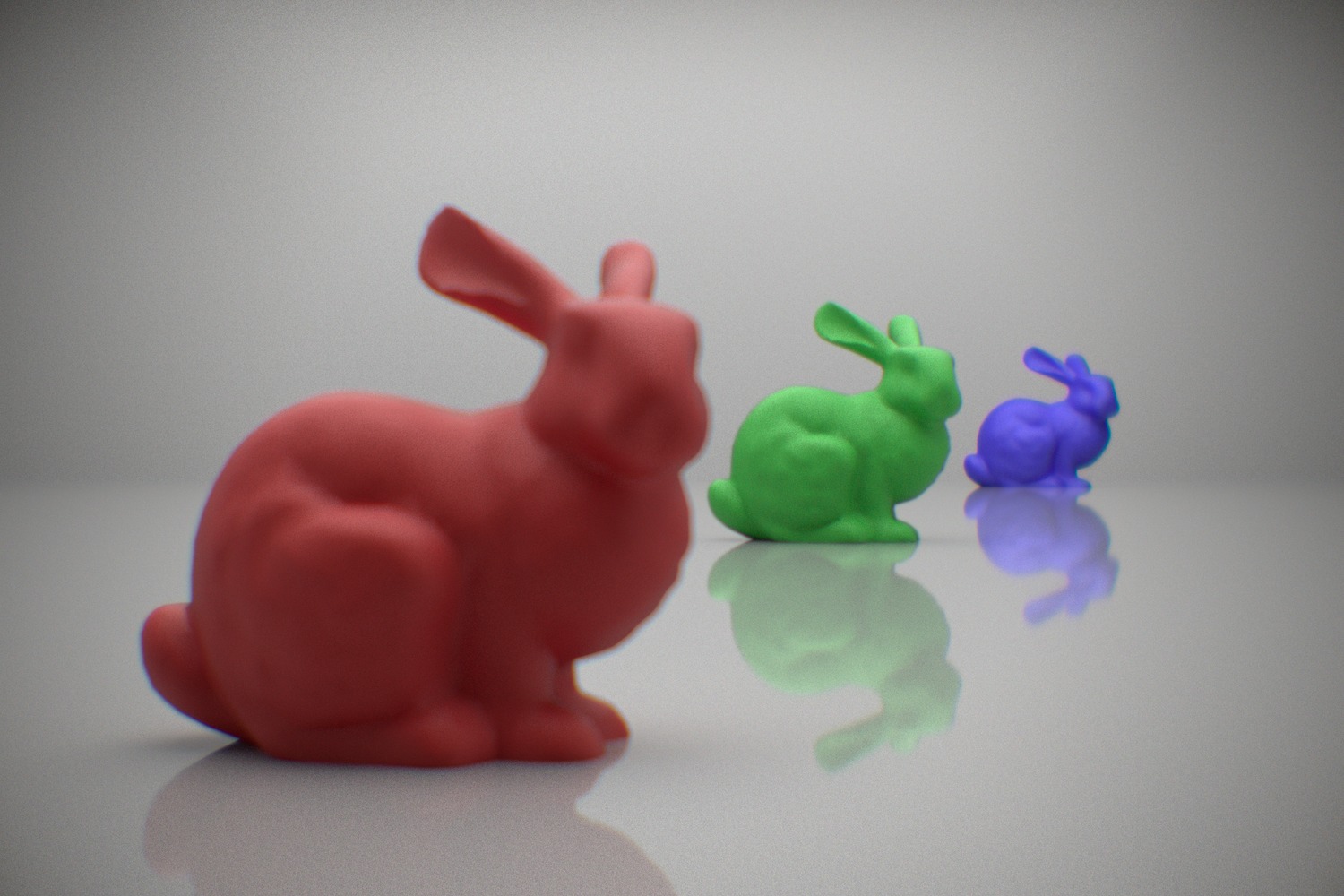} &
        \includegraphics[width=0.3\linewidth]{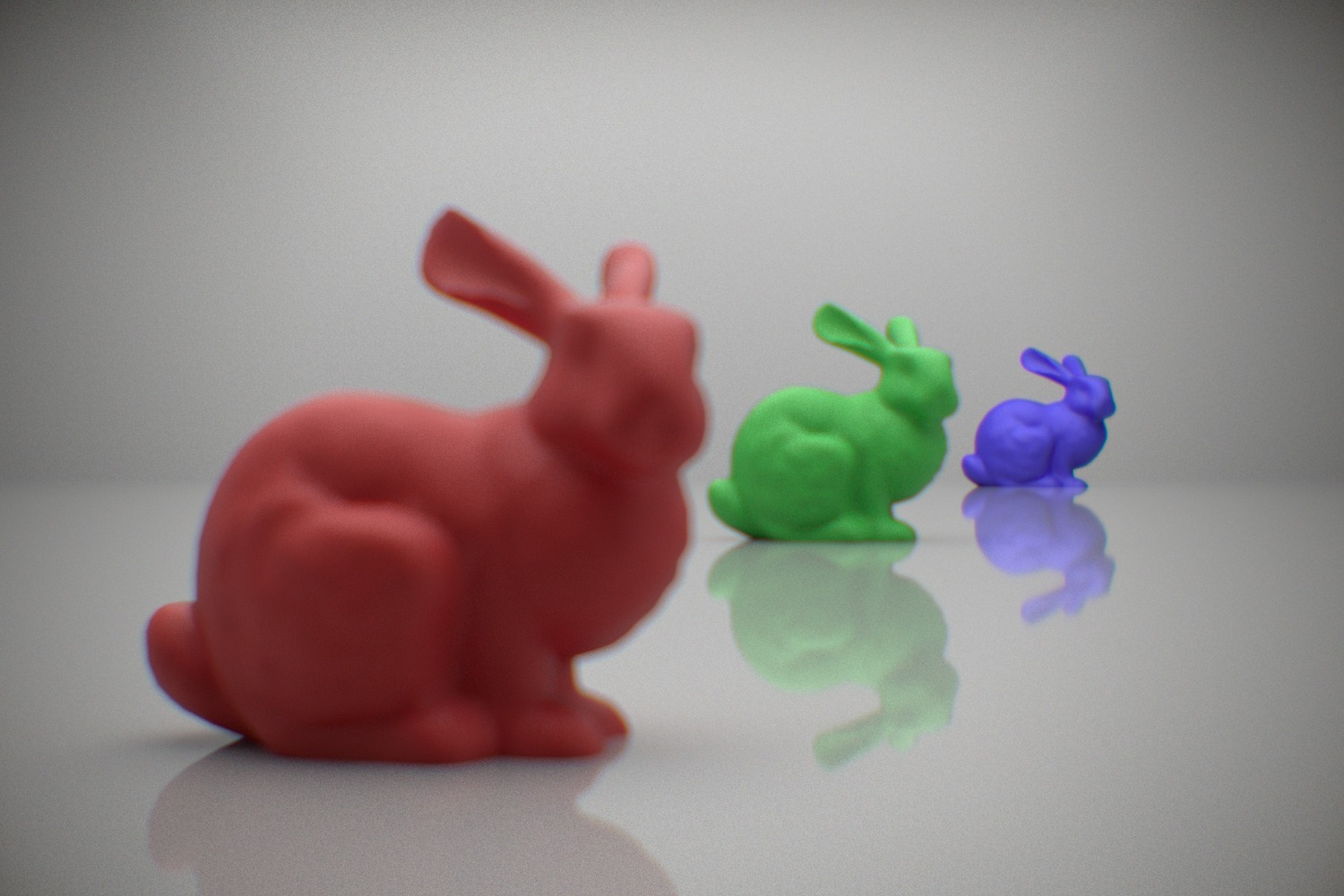} \\
        39.25\,mm & 38.25\,mm & 37.25\,mm
    \end{tabular}
    \caption{
    Demonstration of focus variation by adjusting the sensor position. Each image shows the rendered result at a different CMOS (sensor) distance from the lens's rear pupil, as indicated below each image (in millimeters). The CMOS width is fixed at 24\,mm.
    }
    \label{fig:focus}
\end{figure}

We can also control image sharpness by manually adjusting the CMOS sensor position relative to the last lens element. This process, analogous to focusing in a real camera, involves fine-tuning the distance between the sensor and the lens’s rear pupil. As shown in Figure~\ref{fig:focus}, varying the CMOS depth produces distinct focal planes, allowing us to directly observe the impact of sensor placement on image focus and clarity. 
We used the same precomputed model and rendered all images just by moving the sensor back and forth, demonstrating the accuracy of our model.

\subsubsection*{Rendering Performance}
\begin{figure*}
  \centering
  \setlength{\tabcolsep}{0.2pt}
  \begin{tabular}{c@{}ccc}
    & Wide-angle 22mm lens. Nakamura. & 24mm lens. Canon & 59mm lens. Optical Designer \\
    &
    \begin{minipage}{0.33\linewidth}
        \centering
      \includegraphics[width=0.6\linewidth]{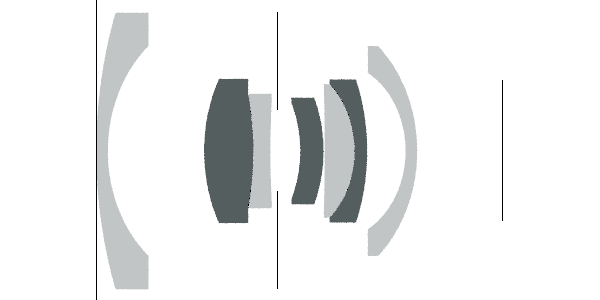}
    \end{minipage} &
    \begin{minipage}{0.33\linewidth}
        \centering
      \includegraphics[width=0.6\linewidth]{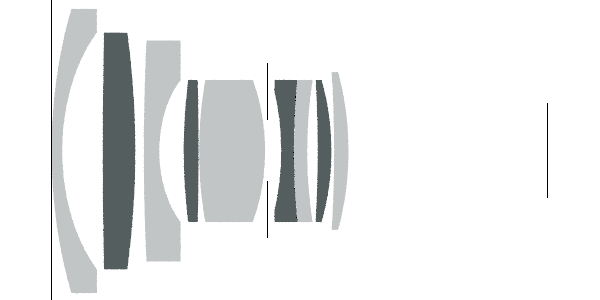}
    \end{minipage} &
    \begin{minipage}{0.33\linewidth}
         \centering
      \includegraphics[width=0.6\linewidth]{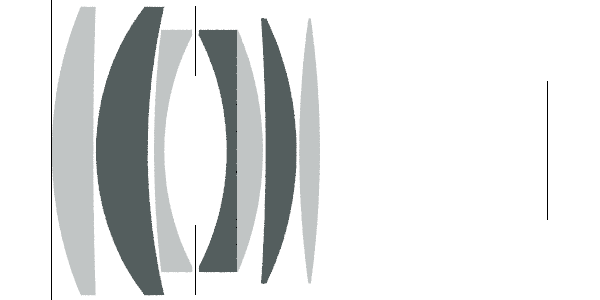}
    \end{minipage} \\
    \rotatebox[origin=c]{90}{Scene "Camera" Ground Truth}&
    \begin{minipage}{0.33\linewidth}
      \includegraphics[width=\linewidth]{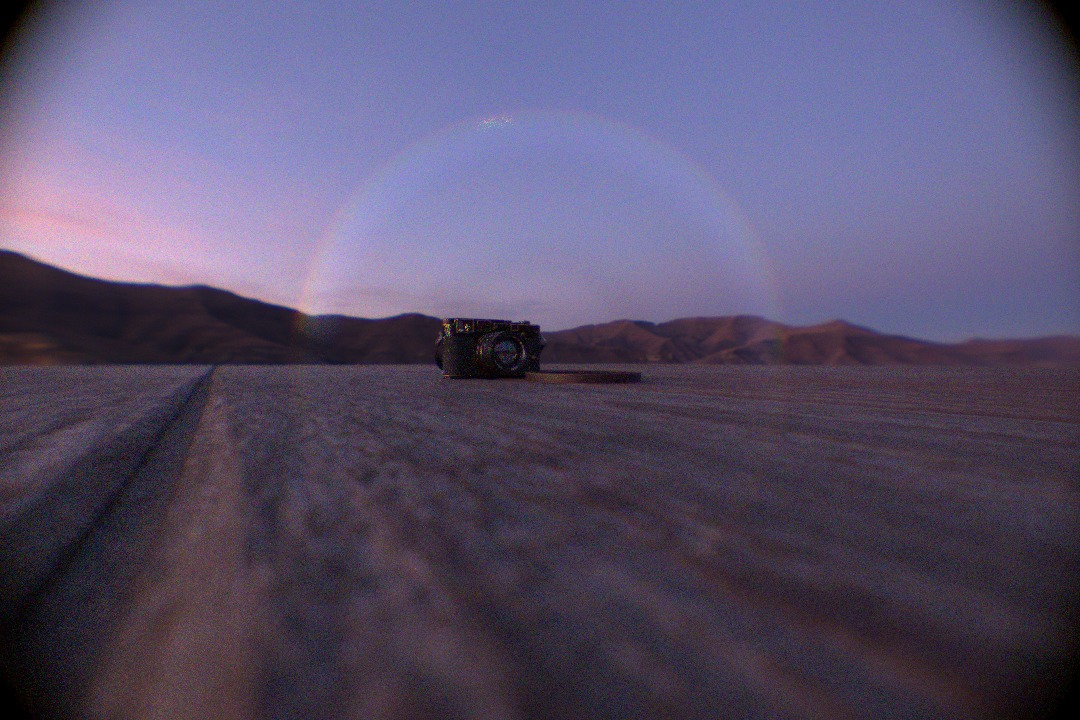}
    \end{minipage} &
    \begin{minipage}{0.33\linewidth}
      \includegraphics[width=\linewidth]{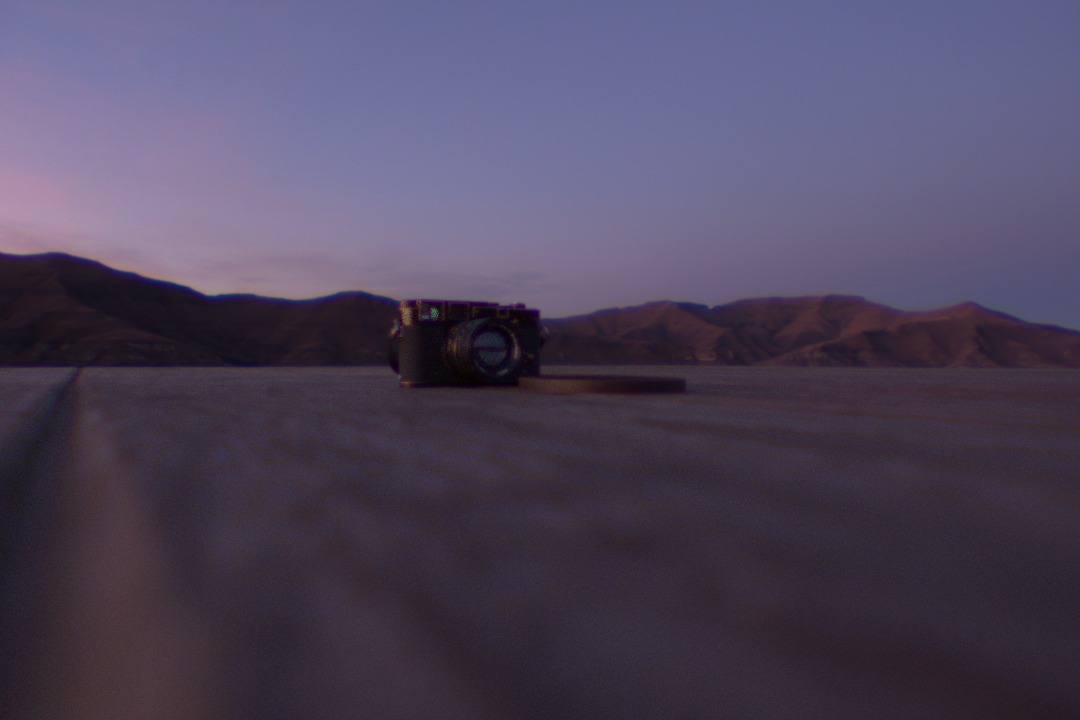}
    \end{minipage} &
    \begin{minipage}{0.33\linewidth}
      \includegraphics[width=\linewidth]{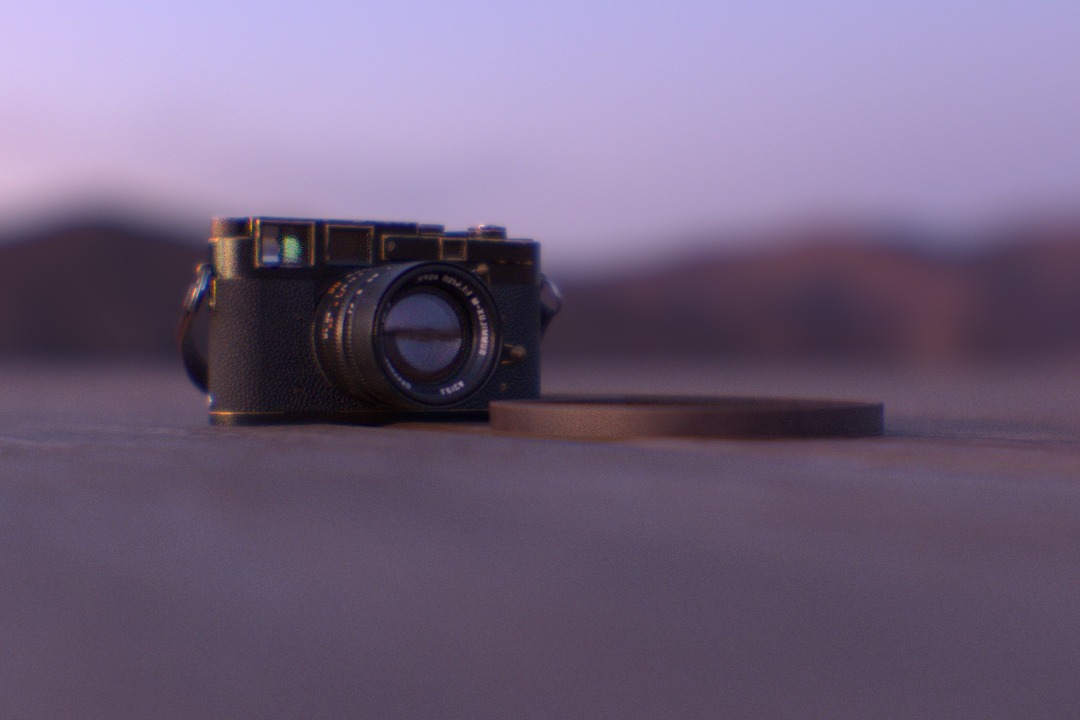}
    \end{minipage} \\
    Time & $\approx$1900s &  $\approx$2300s &  $\approx$2200s \\
    &&&\\
     \newline
      \newline
    \rotatebox[origin=c]{90}{Scene "Camera" Ours}&
    \begin{minipage}{0.33\linewidth}
      \includegraphics[width=\linewidth]{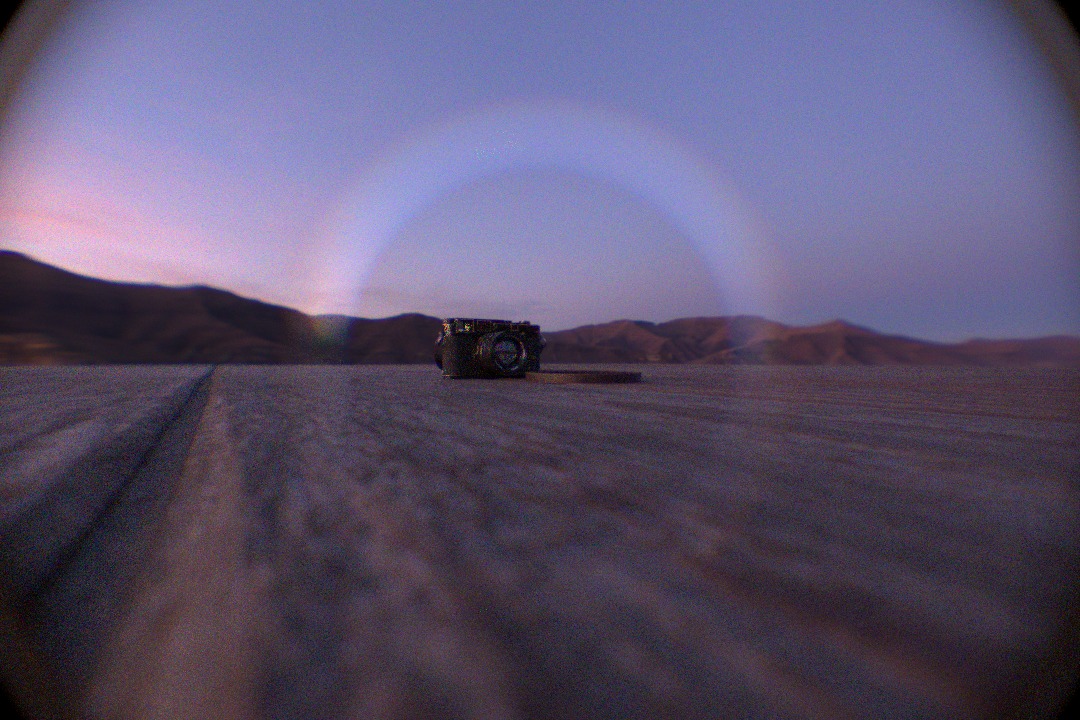}
    \end{minipage} &
    \begin{minipage}{0.33\linewidth}
      \includegraphics[width=\linewidth]{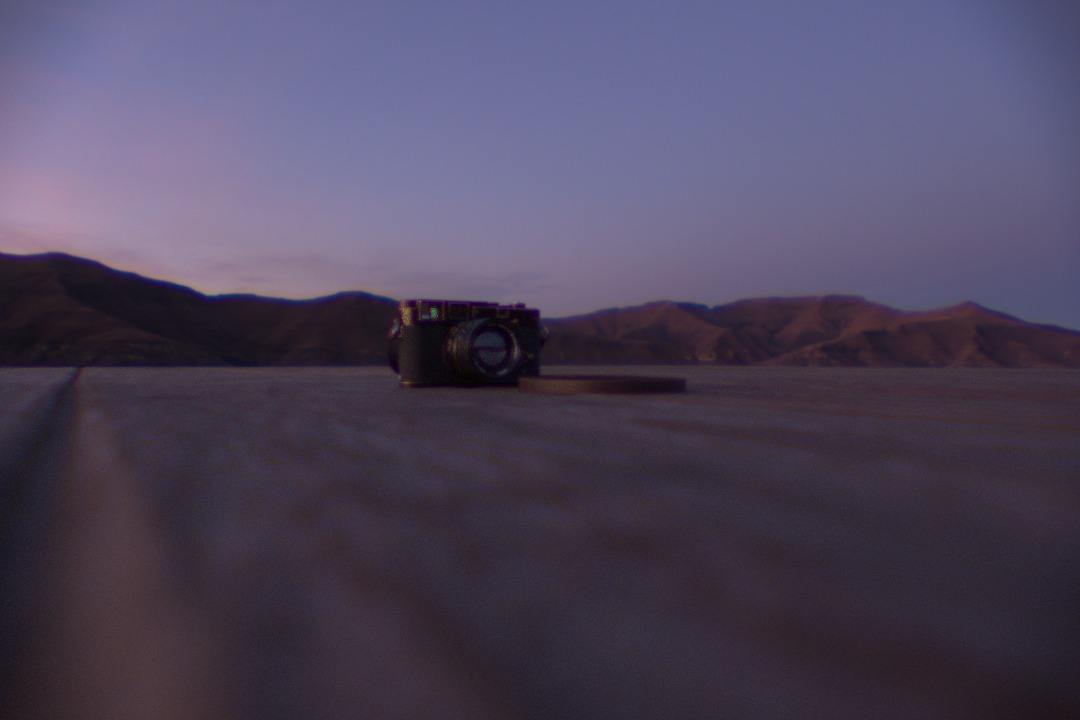}
    \end{minipage} &
    \begin{minipage}{0.33\linewidth}
      \includegraphics[width=\linewidth]{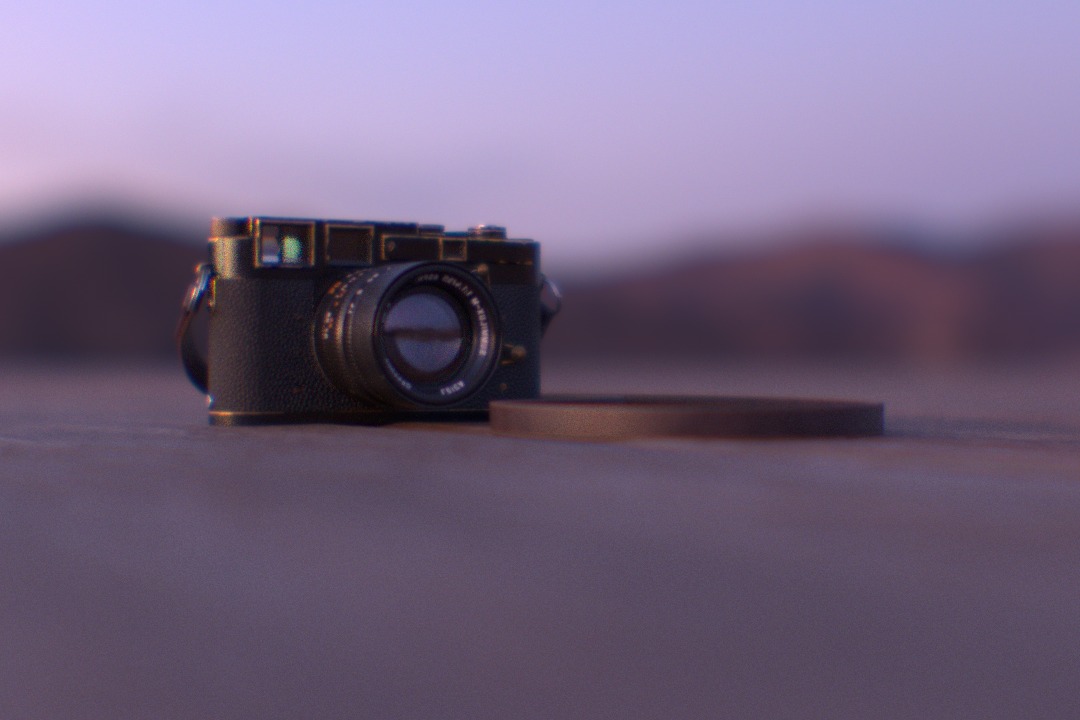}
    \end{minipage} \\
    Time & 230.3s & 246.5s & 250.5s \\
    MAPE& 0.084 & 0.037 & 0.050 \\
    &&&\\
    \rotatebox[origin=c]{90}{Sene "Monk" Ground Truth}&
    \begin{minipage}{0.33\linewidth}
      \includegraphics[width=\linewidth]{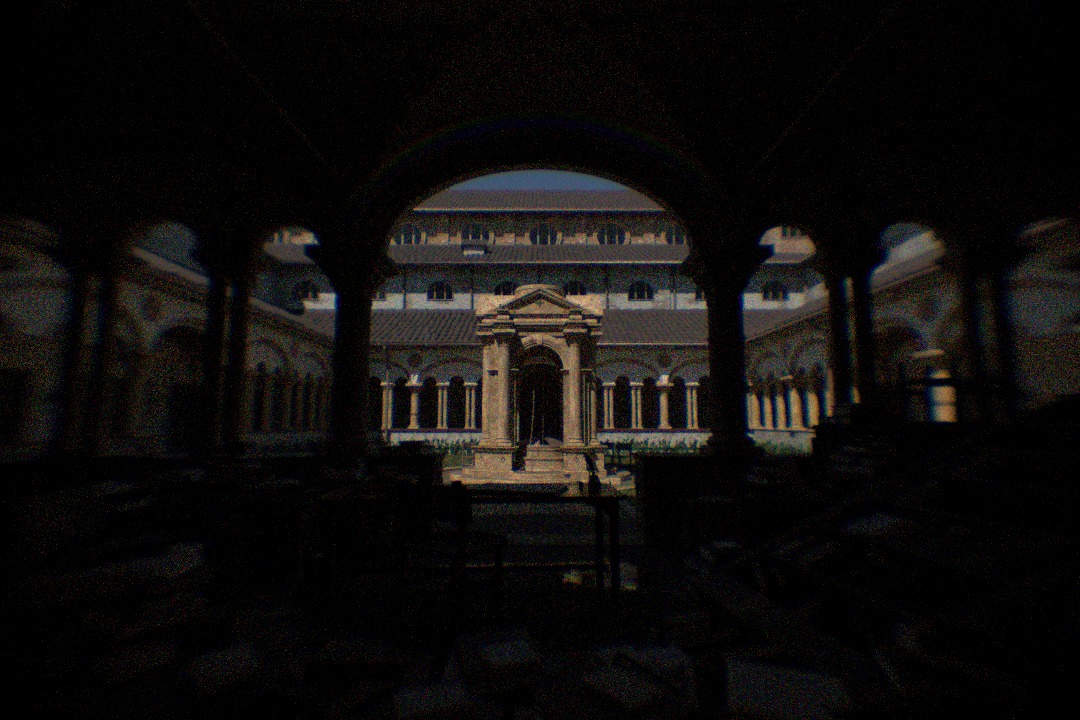}
    \end{minipage} & 
    \begin{minipage}{0.33\linewidth}
      \includegraphics[width=\linewidth]{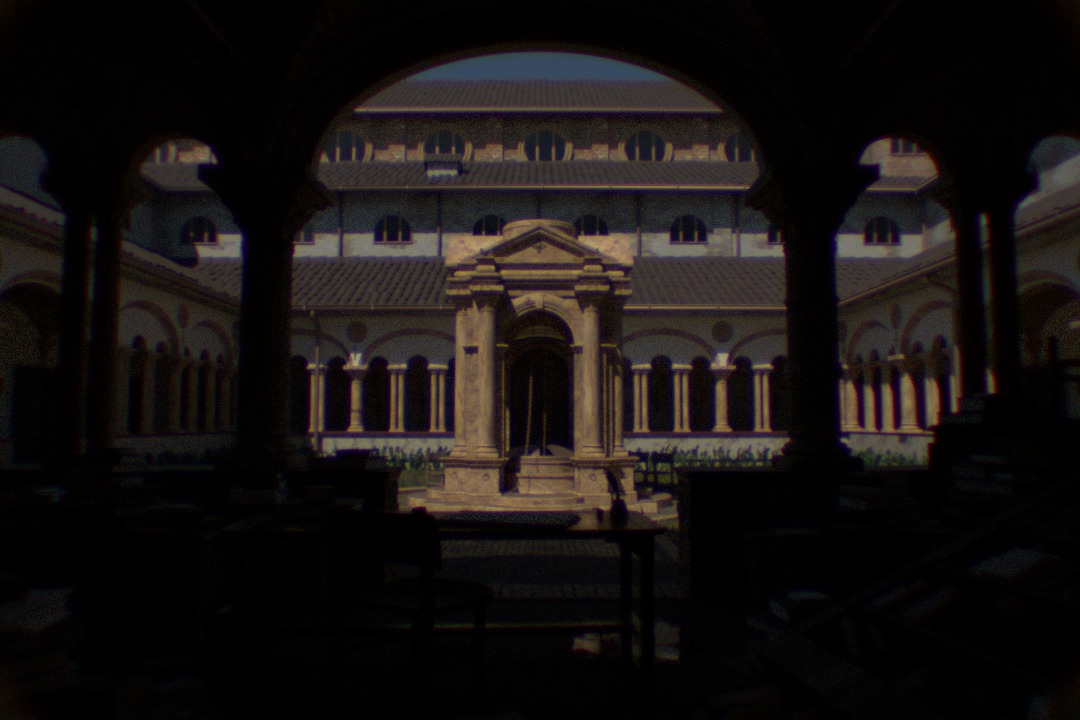}
    \end{minipage} &  
    \begin{minipage}{0.33\linewidth}
      \includegraphics[width=\linewidth]{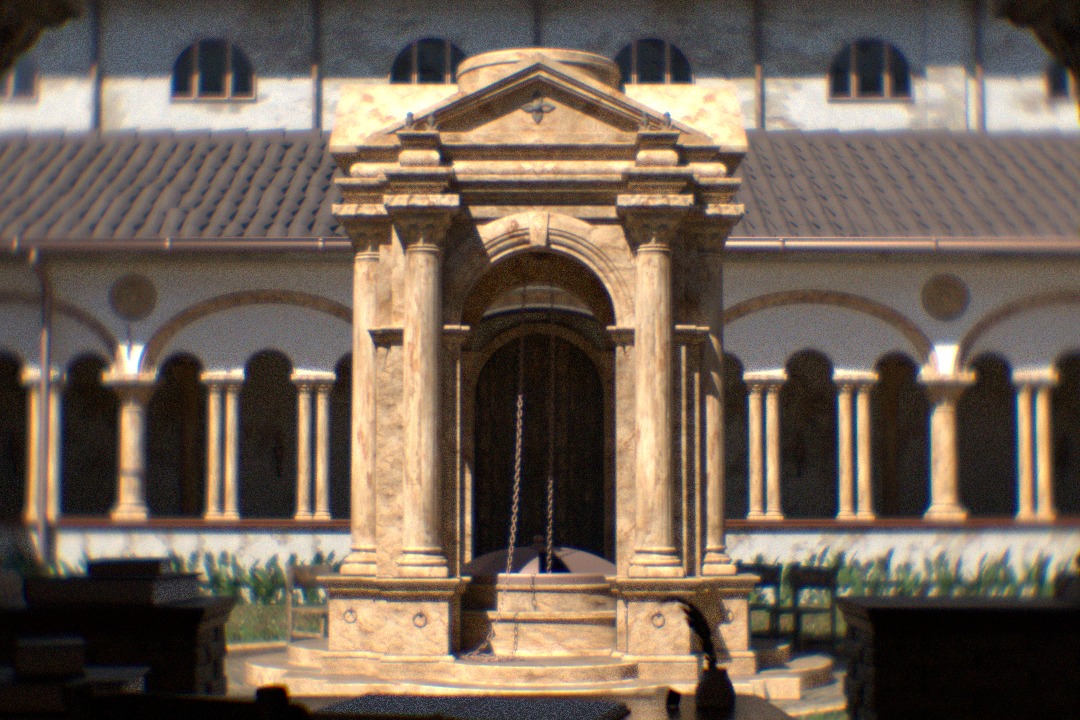}
    \end{minipage} \\ 
    Time & 2101.6s & 3835.2s & 3100.1s \\
    &&&\\
    \rotatebox[origin=c]{90}{Scene "Monk" Ours}&
    \begin{minipage}{0.33\linewidth}
      \includegraphics[width=\linewidth]{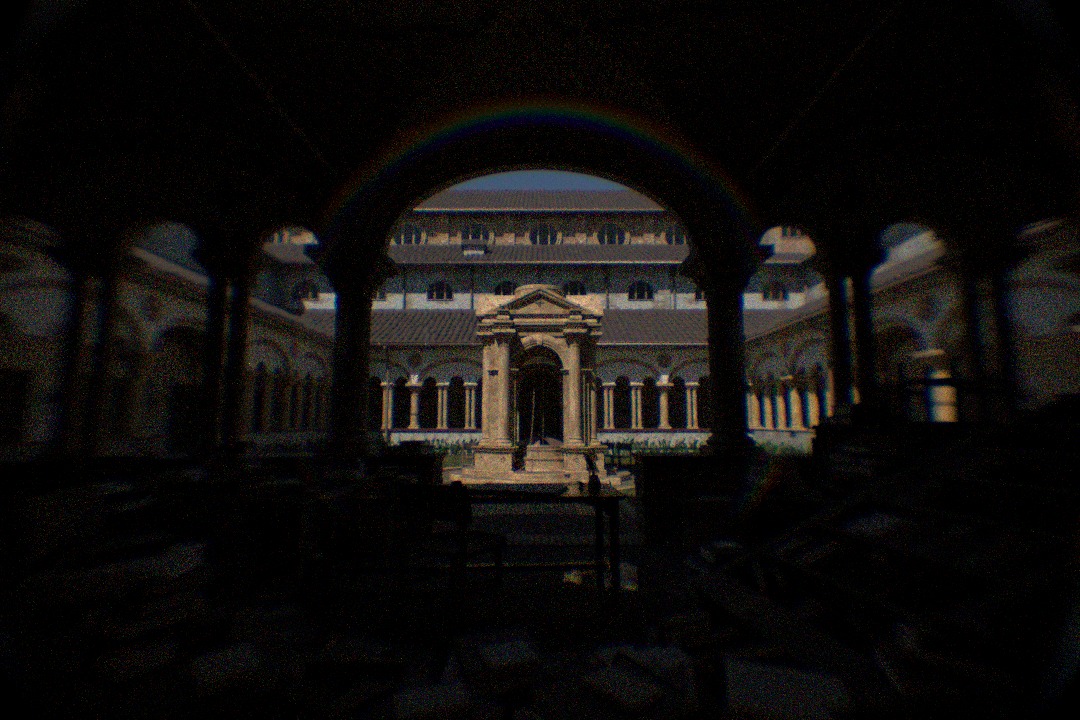}
    \end{minipage} & 
    \begin{minipage}{0.33\linewidth}
      \includegraphics[width=\linewidth]{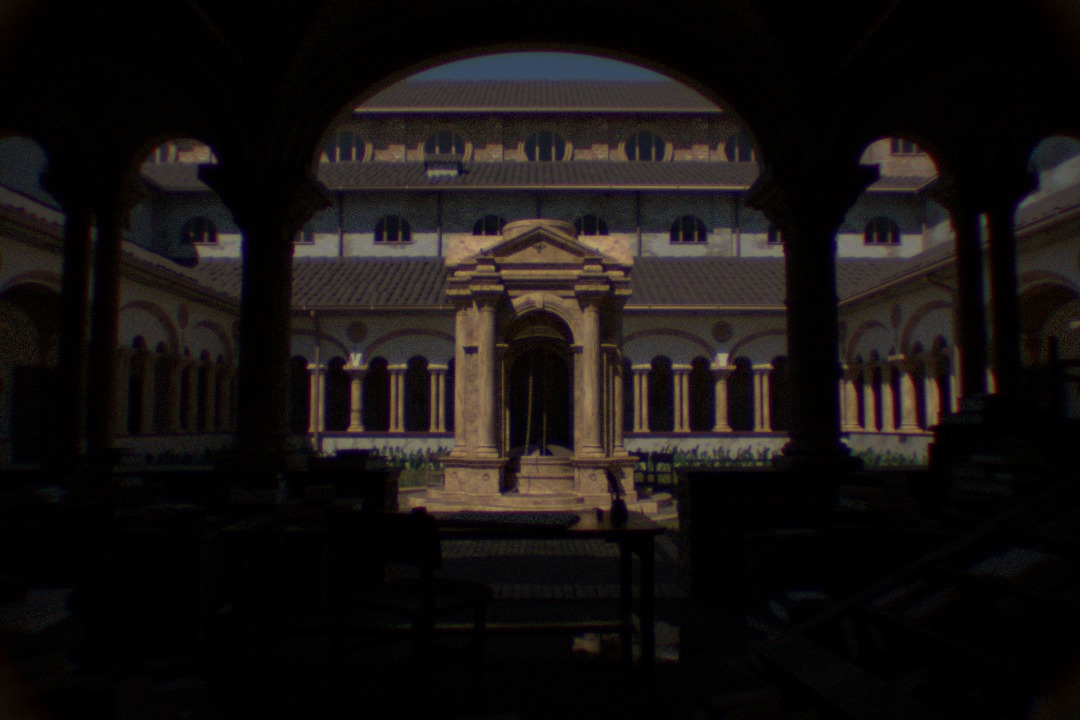}
    \end{minipage} &  
    \begin{minipage}{0.33\linewidth}
      \includegraphics[width=\linewidth]{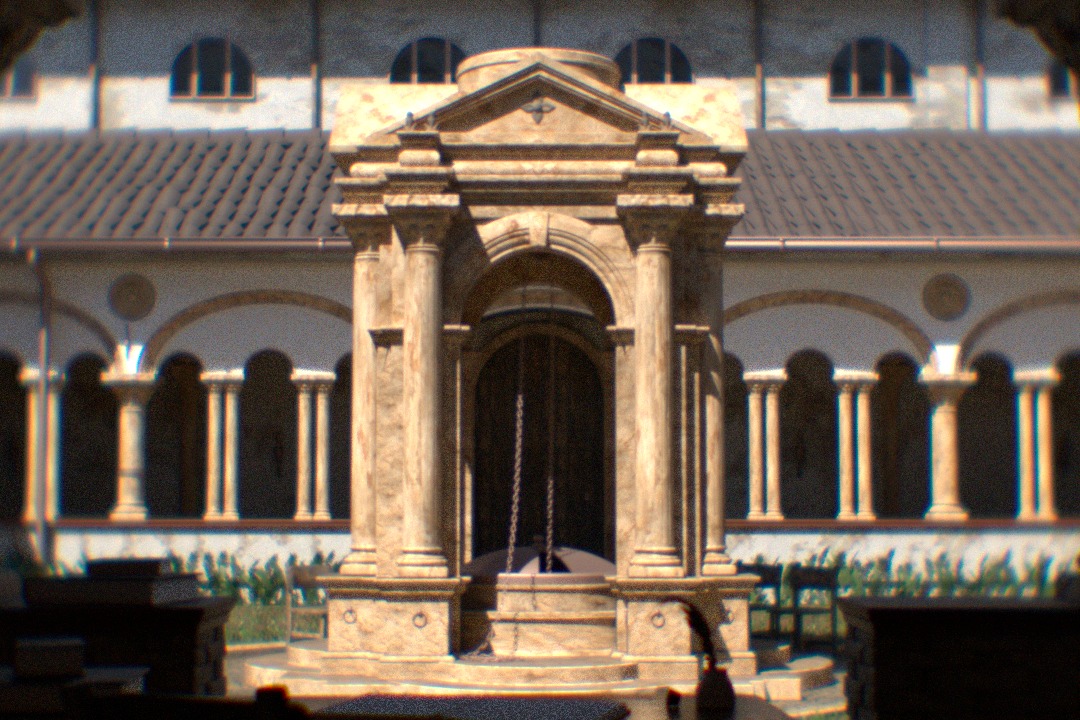}
    \end{minipage} \\ 
    Time & 141.2s & 197.2s & 258.3s \\
    MAPE& 0.082 & 0.112 & 0.142 
  \end{tabular}
  \caption{Equal sample comparison between ray traced lens systems and our neural lens transport.}
  \label{fig:path-tracing}
\end{figure*}

We further evaluate the efficiency of our approach by comparing the rendering performance of traditional path tracing with a ray-traced lens system to our neural lens transport method (see Figure~\ref{fig:path-tracing}). Our method not only produces images that are visually comparable to the ground truth, but also achieves an order of magnitude speedup of roughly 12$\sim$15x in rendering. This performance gain arises because ray-traced lens systems are less GPU-friendly due to higher register usage and increased control flow divergence. In contrast, our fused MLP can be implemented efficiently on the GPU, resulting in significant performance improvements across various scenes and lens configurations.

\subsubsection*{Importance of Classifier Network}
\begin{figure}
    \centering
    \begin{tabular}{c c}
        \begin{minipage}[b]{0.49\linewidth}
            \centering
            \includegraphics[width=\linewidth]{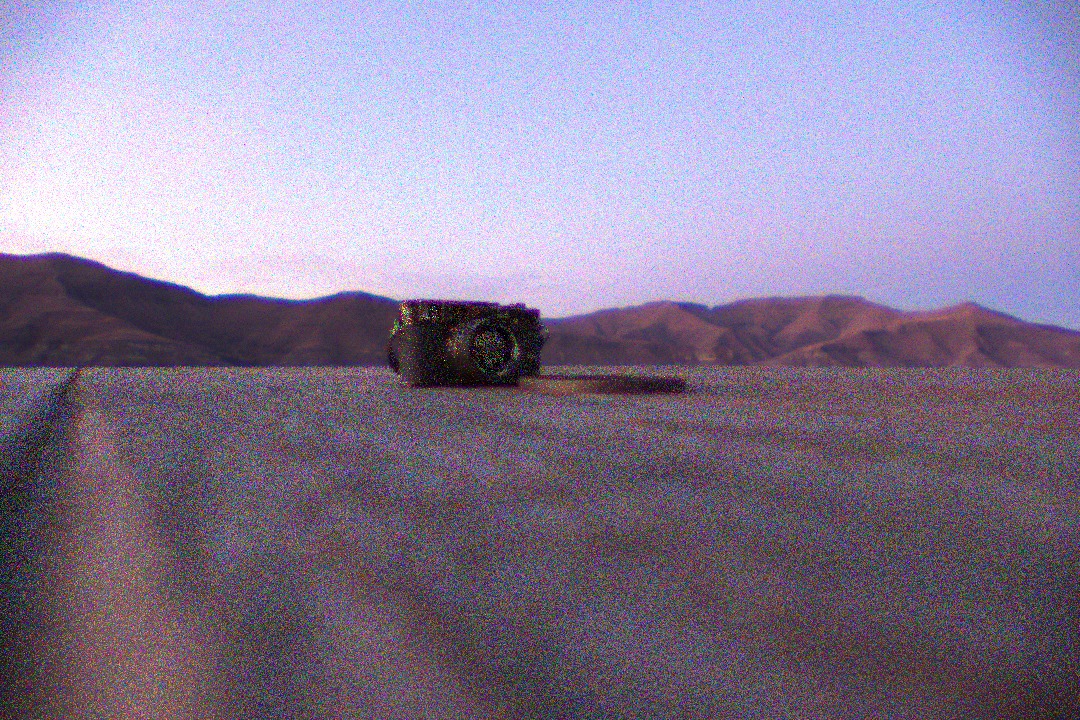}
        \end{minipage}&
        \hfill
        \begin{minipage}[b]{0.49\linewidth}
            \centering
            \includegraphics[width=\linewidth]{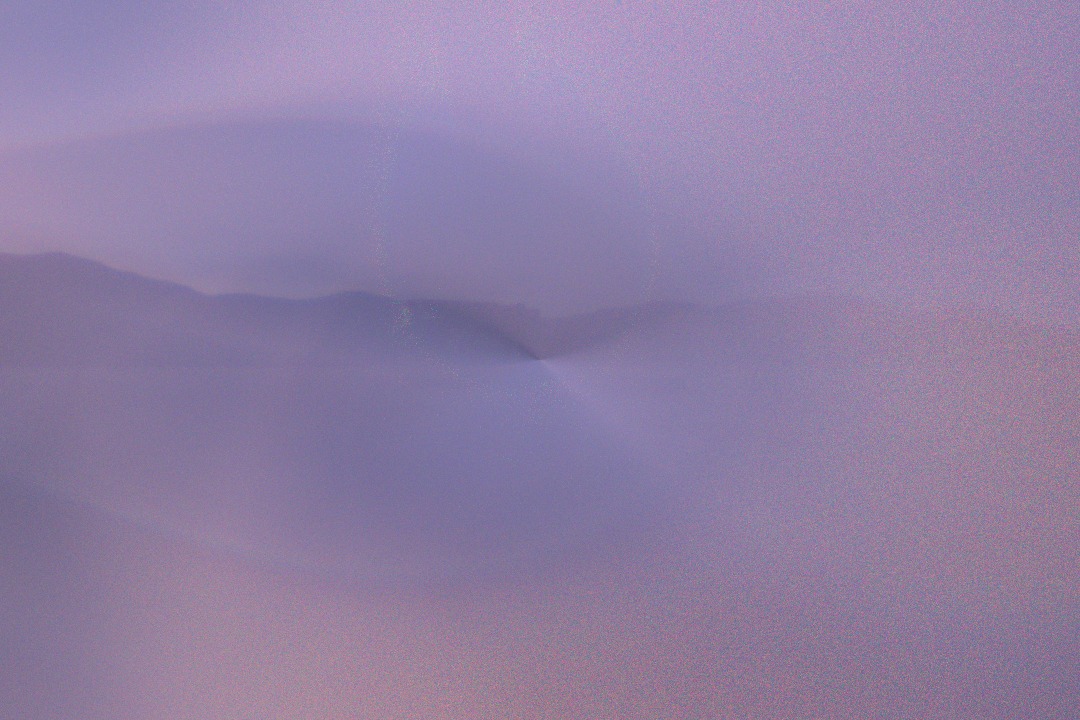}
        \end{minipage}
    \end{tabular}
    \caption{We compared the image renderer with (left) and without our classifier (right, -5.0 exposure). The classifier plays a crucial rule in eliminating invalid rays. Without the classifier, all rays are assumed to be valid, including those hitting the aperture and barrel, eventually resulting in an overly-exposed, incorrect image.}
    \label{fig:mask}
\end{figure}

To demonstrate the necessity of using the classifier to handle discontinuities, we disable the classifier in Figure~\ref{fig:mask}. Without the classifier, all rays, including those that are occluded, contributed to the final image, resulting in over exposed and a heavily blurred image.

\section{Limitations and Future Work}
We presented a robust and efficient method for precomputed light transport within lens system.
Our neural lens transport map handles discontinuities with the classifier-regressor architecture, employs a tanh-MLP to ensure the predicted exiting ray is $C^1$ continuous. The overall pipeline demonstrates improved accuracy over the polynomial model and accelerated rendering performance.
Our current implementation focuses on symmetric lenses and apertures. Generalizing our method to support asymmetric lens system would further enhance the versatility of our method.
Our method currently only handles static lens system that has a fixed focal length and aperture without retrain the neural model. It is possible to extend our method to support zoom lens by using more than one MLP as well as dynamic aperture sizes. 
Our precomputation currently assumes geometric optics, but adding wave optical effects such as diffraction to our model is a challenging future work. 

\begin{acks}
This work originated as a fourth-month research project for the CS 888 course (Winter 2025) at the University of Waterloo, details of which can be found at \url{https://cs.uwaterloo.ca/~thachisu/CS888_W25/}. We would also like to thank Ryan Zhu, Hongfei Huang, and Ege Ciklabakkal for their feedback and camaraderie throughout the initial development of this work.
\end{acks}

\bibliographystyle{ACM-Reference-Format}
\bibliography{main}

@string{TOG = "ACM Trans. Graph."}

@string{CGF = "Comput. Graph. Forum"}

@inproceedings{schrade2016sparse,
  title={Sparse high-degree polynomials for wide-angle lenses},
  author={Schrade, Emanuel and Hanika, Johannes and Dachsbacher, Carsten},
  booktitle={Computer Graphics Forum},
  volume={35},
  number={4},
  pages={89--97},
  year={2016},
  organization={Wiley Online Library}
}

@article{zheng2017adaptive,
  title={Adaptive sparse polynomial regression for camera lens simulation},
  author={Zheng, Quan and Zheng, Changwen},
  journal={The Visual Computer},
  volume={33},
  pages={715--724},
  year={2017},
  publisher={Springer}
}

@inproceedings{pekkarinen2019physically,
  title={Physically based lens flare rendering in" The Lego Movie 2"},
  author={Pekkarinen, Erik and Balzer, Michael},
  booktitle={Proceedings of the 2019 Digital Production Symposium},
  pages={1--3},
  year={2019}
}

@inproceedings{teh2024aperture,
  title={Aperture-Aware Lens Design},
  author={Teh, Arjun and Gkioulekas, Ioannis and O'Toole, Matthew},
  booktitle={ACM SIGGRAPH 2024 Conference Papers},
  pages={1--10},
  year={2024}
}

@article{tseng2021differentiable,
  title={Differentiable compound optics and processing pipeline optimization for end-to-end camera design},
  author={Tseng, Ethan and Mosleh, Ali and Mannan, Fahim and St-Arnaud, Karl and Sharma, Avinash and Peng, Yifan and Braun, Alexander and Nowrouzezahrai, Derek and Lalonde, Jean-Francois and Heide, Felix},
  journal={ACM Transactions on Graphics (TOG)},
  volume={40},
  number={2},
  pages={1--19},
  year={2021},
  publisher={ACM New York, NY}
}

@article{rossmann1969point,
  title={Point spread-function, line spread-function, and modulation transfer function: tools for the study of imaging systems},
  author={Rossmann, Kurt},
  journal={Radiology},
  volume={93},
  number={2},
  pages={257--272},
  year={1969},
  publisher={The Radiological Society of North America}
}

@article{bodonyi2024real,
  title={Real-time ray transfer for lens flare rendering using sparse polynomials},
  author={Bodonyi, Andrea and Csoba, Istv{\'a}n and Kunkli, Roland},
  journal={The Visual Computer},
  pages={1--18},
  year={2024},
  publisher={Springer}
}

@inproceedings{lee2013practical,
  title={Practical real-time lens-flare rendering},
  author={Lee, Sungkil and Eisemann, Elmar},
  booktitle={Computer Graphics Forum},
  volume={32},
  number={4},
  pages={1--6},
  year={2013},
  organization={Wiley Online Library}
}

@misc{openopticaldesigner,
  author       = {Alex Bock},
  title        = {Open Optical Designer},
  year         = 2023,
  howpublished = {\url{https://github.com/alexbock/open-optical-designer}},
  note         = {Accessed: 2025-05-22}
}

@article{Zheng2022LuisaRender,
    author = {Zheng, Shaokun and Zhou, Zhiqian and Chen, Xin and Yan, Difei and Zhang, Chuyan and Geng, Yuefeng and Gu, Yan and Xu, Kun},
    title = {LuisaRender: A High-Performance Rendering Framework with Layered and Unified Interfaces on Stream Architectures},
    year = {2022},
    issue_date = {December 2022},
    publisher = {Association for Computing Machinery},
    address = {New York, NY, USA},
    volume = {41},
    number = {6},
    issn = {0730-0301},
    url = {https://doi.org/10.1145/3550454.3555463},
    doi = {10.1145/3550454.3555463},
    journal = {ACM Trans. Graph.},
    month = {nov},
    articleno = {232},
    numpages = {19}
}

@article{Steinert2011LensSimulation,
author = {Steinert, B. and Dammertz, H. and Hanika, J. and Lensch, H. P. A.},
title = {General Spectral Camera Lens Simulation},
journal = {Computer Graphics Forum},
volume = {30},
number = {6},
pages = {1643-1654},
doi = {https://doi.org/10.1111/j.1467-8659.2011.01851.x},
url = {https://onlinelibrary.wiley.com/doi/abs/10.1111/j.1467-8659.2011.01851.x},
eprint = {https://onlinelibrary.wiley.com/doi/pdf/10.1111/j.1467-8659.2011.01851.x},
year = {2011}
}

@article{DBLP:journals/tog/HullinESL11,
  author       = {Matthias B. Hullin and
                  Elmar Eisemann and
                  Hans{-}Peter Seidel and
                  Sungkil Lee},
  title        = {Physically-based real-time lens flare rendering},
  journal      = {{ACM} Trans. Graph.},
  volume       = {30},
  number       = {4},
  pages        = {108},
  year         = {2011},
  url          = {https://doi.org/10.1145/2010324.1965003},
  doi          = {10.1145/2010324.1965003},
  bibsource    = {dblp computer science bibliography, https://dblp.org}
}

@article{DBLP:journals/cgf/HullinHH12,
  author       = {Matthias B. Hullin and
                  Johannes Hanika and
                  Wolfgang Heidrich},
  title        = {Polynomial Optics: {A} Construction Kit for Efficient Ray-Tracing
                  of Lens Systems},
  journal      = {Comput. Graph. Forum},
  volume       = {31},
  number       = {4},
  pages        = {1375--1383},
  year         = {2012},
  url          = {https://doi.org/10.1111/j.1467-8659.2012.03132.x},
  doi          = {10.1111/J.1467-8659.2012.03132.X},
  bibsource    = {dblp computer science bibliography, https://dblp.org}
}

@article{DBLP:journals/cgf/JooKLEL16,
  author       = {Hyuntae Joo and
                  Soonhyeon Kwon and
                  Sangmin Lee and
                  Elmar Eisemann and
                  Sungkil Lee},
  title        = {Efficient Ray Tracing Through Aspheric Lenses and Imperfect Bokeh
                  Synthesis},
  journal      = {Comput. Graph. Forum},
  volume       = {35},
  number       = {4},
  pages        = {99--105},
  year         = {2016},
  url          = {https://doi.org/10.1111/cgf.12953},
  doi          = {10.1111/CGF.12953},
  bibsource    = {dblp computer science bibliography, https://dblp.org}
}

@article{Hanika:2014:Lens,
  title = {Efficient {Monte Carlo} Rendering with Realistic Lenses},
  author = {Johannes Hanika and Carsten Dachsbacher},
  year = 2014,
  volume = 33,
  number = {2},
  journal = {Computer Graphics Forum (Proceedings of Eurographics)},
  month = {April},
  pages = {323--332}
}

@article{DBLP:conf/siggraph/KolbMH95,
  author       = {Craig E. Kolb and
                  Don P. Mitchell and
                  Pat Hanrahan},
  editor       = {Susan G. Mair and
                  Robert Cook},
  title        = {A realistic camera model for computer graphics},
  booktitle    = {Proceedings of the 22nd Annual Conference on Computer Graphics and
                  Interactive Techniques, {SIGGRAPH} 1995, Los Angeles, CA, USA, August
                  6-11, 1995},
  pages        = {317--324},
  publisher    = {{ACM}},
  year         = {1995},
  url          = {https://doi.org/10.1145/218380.218463},
  doi          = {10.1145/218380.218463},
  bibsource    = {dblp computer science bibliography, https://dblp.org}
}

@article{Levenberg:1944:MSC,
    title = {A method for the solution of certain non-linear problems in least squares},
    author = {Levenberg, Kenneth},
    journal = {Quarterly of applied mathematics},
    volume = {2},
    number = {2},
    pages = {164--168},
    year = {1944}
}

@article{Zheng:2017:NDC,
    title = {NeuroLens: Data-Driven Camera Lens Simulation Using Neural Networks},
    author = {Zheng, Quan and Zheng, Changwen},
    booktitle = CGF,
    volume = {36},
    number = {8},
    pages = {390--401},
    year = {2017},
    publisher = {Eurographics Association}
}

@article{hachisuka_robust_2011,
	title = {Robust adaptive photon tracing using photon path visibility},
	volume = {30},
	issn = {0730-0301, 1557-7368},
	url = {https://dl.acm.org/doi/10.1145/2019627.2019633},
	doi = {10.1145/2019627.2019633},
	pages = {1--11},
	number = {5},
	journaltitle = {{ACM} Transactions on Graphics},
	shortjournal = {{ACM} Trans. Graph.},
	author = {Hachisuka, Toshiya and Jensen, Henrik Wann},
	urldate = {2023-11-19},
    year = {2011},
	date = {2011-10},
	langid = {english},
}

@article{Zheng:2017:NeuroLens,
    author = {Zheng, Quan and Zheng, Changwen},
    title = {NeuroLens: Data-Driven Camera Lens Simulation Using Neural Networks},
    journal = {Computer Graphics Forum},
    volume = {36},
    number = {8},
    pages = {390-401},
    doi = {https://doi.org/10.1111/cgf.13087},
    url = {https://onlinelibrary.wiley.com/doi/abs/10.1111/cgf.13087},
    eprint = {https://onlinelibrary.wiley.com/doi/pdf/10.1111/cgf.13087},
    year = {2017}
}

\end{document}